\documentclass{article}
\pdfoutput=1
\usepackage{jheppub}
\notoc

\usepackage{amssymb}
\usepackage{tikz}
\usepackage{amsmath}
\usepackage{hyperref}

\usepackage[numbers,sort&compress]{natbib} 

\newcommand{\ka}{\kappa}

\newcommand{\de}{\delta} 

\newcommand{\ba}{\begin{align}}
\newcommand{\ea}{\end{align}}	
\newcommand{\eref}[1]{Eq.~(\ref{#1})}
\newcommand{\fref}[1]{Fig.~\ref{#1}}
\newcommand{\nnnl}{\nonumber\\}	

\newcommand{\beq}{\begin{eqnarray}}
\newcommand{\eeq}{\end{eqnarray}}

\title{On the influence of three-point functions on the propagators of Landau gauge Yang-Mills theory}

\author[a]{Markus Q. Huber,}

\author[a]{Lorenz von Smekal}

\affiliation[a]{Institut f\"ur Kernphysik, Technische Universit\"at Darmstadt, Schlossgartenstr. 2, 64289 Darmstadt, Germany}

\emailAdd{markus.huber@physik.tu-darmstadt.de}
\emailAdd{lorenz.smekal@physik.tu-darmstadt.de}

\abstract{We solve the Dyson-Schwinger equations of the ghost and gluon propagators of Landau gauge Yang-Mills theory together with that of the ghost-gluon vertex. The latter plays a central role in many truncation schemes for functional equations. By including it dynamically we can determine its influence on the propagators. We also suggest a new model for the three-gluon vertex motivated by lattice data which plays a crucial role to obtain stable solutions when the ghost-gluon vertex is included. We find that both vertices have a sizable quantitative impact on the mid-momentum regime and contribute to the reduction of the gap between lattice and Dyson-Schwinger equation results. Furthermore, we establish that the three-gluon vertex dressing turns negative at low momenta as suggested by lattice results in three dimensions.
}


\begin{document}
\maketitle

\section{Introduction}

The propagators of Yang-Mills theory have been widely investigated because, on one hand, they provide the starting point for calculations of hadronic observables and investigations of the phase diagram of strongly interacting matter as described by quantum chromodynamics (QCD). On the other hand, they provide some direct insight into the mechanisms behind non-perturbative phenomena like confinement and dynamical chiral symmetry breaking.
Typically, in the Landau gauge the ghost-gluon vertex has been a pivotal object in such studies when using functional methods, since its simple structure is the basis for most truncation schemes. The original motivation for this was Taylor's non-renormalization argument for the Landau gauge \cite{Taylor:1971ff}. The employed truncation schemes were so successful that the complications arising by including the vertex dynamically into numeric calculations have been postponed for some time. In the meantime additional information was gathered \cite{Lerche:2002ep,Schleifenbaum:2004id,Schleifenbaum:2006bq,Alkofer:2008jy,Alkofer:2008dt,Fischer:2009tn,Boucaud:2011eh,Dudal:2012zx} with functional methods backed up by calculations on the lattice \cite{Cucchieri:2004sq,Cucchieri:2006tf,Sternbeck:2006rd,Cucchieri:2008qm} that solidified the reliability of the used truncation schemes and supported the expectations that the modifications induced by a self-consistent inclusion of the vertex are small and only on the quantitative level. Here we test these expectations by including the ghost-gluon vertex dynamically into the system of equations we solve, in order to assess any quantitative or even qualitative differences.

While major qualitative changes are certainly not expected, the extent to which quantitative modifications occur is not clear. At least some effects are anticipated, as comparisons of calculations with different vertex models show \cite{Pennington:2011xs}. Since the ghost-gluon vertex is an important part of the Yang-Mills sector, we should study it as well as possible to ensure that we fully understand the non-perturbative infrared regime of Landau gauge QCD. Furthermore, it recently turned out that a simple ghost-gluon vertex as employed in the vacuum is insufficient for non-zero temperature calculations \cite{Fister:2011uw}. Finally, it is desirable to improve existing results for the propagators quantitatively in order to become more competitive with lattice results. The inclusion of the ghost-gluon vertex constitutes one step in this direction, also because it is a prerequisite for the investigation of other quantities like the gluonic vertices. The lowest one, the three-gluon vertex, is expected to have even more quantitative influence on the propagators. In order to test this assumption, we investigate a new three-gluon vertex model. It is Bose symmetric, has the correct ultraviolet (UV) anomalous dimension and features a zero crossing as observed in three-dimensional lattice data. The latter property has not been clearly observed in four dimensions yet. Thus we also compute the infrared (IR) leading contribution of the three-gluon vertex for one momentum configuration to confirm this.

Perturbatively the ghost-gluon vertex is well studied. A detailed one-loop calculation for general external momenta can be found in Ref. \cite{Davydychev:1996pb} and a three-loop calculation for the so-called asymmetric point in Ref. \cite{Chetyrkin:2000dq}. An often used argument about its IR behavior goes back to Taylor \cite{Taylor:1971ff}. We will comment on it in more detail below in Sec.~\ref{sec:ghg}. Further analytic studies have been performed with Dyson-Schwinger equations (DSEs) \cite{Alkofer:2008jy,Alkofer:2008dt} and first functional renormalization group calculations for non-zero temperature were also done \cite{Fister:2011uw}.
An operator product expansion analysis can be found in Ref.~\cite{Boucaud:2011eh}. It was subsequently used for an indirect determination of the parameters by studying its effects in the ghost DSE \cite{Dudal:2012zx}.
On the lattice the vertex was investigated in Refs.~\cite{Cucchieri:2004sq,Cucchieri:2006tf,Sternbeck:2006rd,Cucchieri:2008qm}, but unfortunately the results are by far not as detailed as for the propagators. Still, they are the most reliable source of quantitative information. Semi-perturbative results from DSEs can be found in \cite{Schleifenbaum:2004id, Schleifenbaum:2004dt}. Here we will present the first calculation with full momentum dependence and back coupling effects.

The dynamical inclusion of the ghost-gluon vertex required at the same time to improve on the three-gluon vertex model. The reason is that in the gluon propagator DSE the contributions of the ghost and gluon loops naturally balance to keep the gluon dressing function positive. If we now change the ghost-gluon vertex this balance might get lost and the gluon loop could take over in the mid-momentum regime rendering the gluon propagator negative. This gives a natural constraint on any vertex model and was one motivation for introducing the new model for the three-gluon vertex that takes into account the known properties from lattice \cite{Cucchieri:2008qm} and Dyson-Schwinger studies \cite{Alkofer:2004it,Alkofer:2008dt}. The improved vertex has a considerable effect in the mid-momentum regime, where the difference to lattice calculations is most evident.

We start with an overview of the Green functions employed here and their DSEs in Sec.~\ref{sec:DSEs}. There also some remarks about the Taylor theorem can be found and the model for the three-gluon vertex is explained. The employed methods are reviewed in Sec.~\ref{sec:methods}. Results are presented in Secs.~\ref{sec:results_3g} and \ref{sec:results_ghg}. We conclude with a summary in Sec.~\ref{sec:summary}. In two appendices we present the kernels of the DSEs and further details on the UV behavior of the propagators. In Ref.~\cite{Huber:2013xb} a short summary of this article can be found.

\section{The system of Dyson-Schwinger equations}
\label{sec:DSEs}

\subsection{The two-point Dyson-Schwinger equations}

\begin{figure}[tb]
\includegraphics[width=8.3cm]{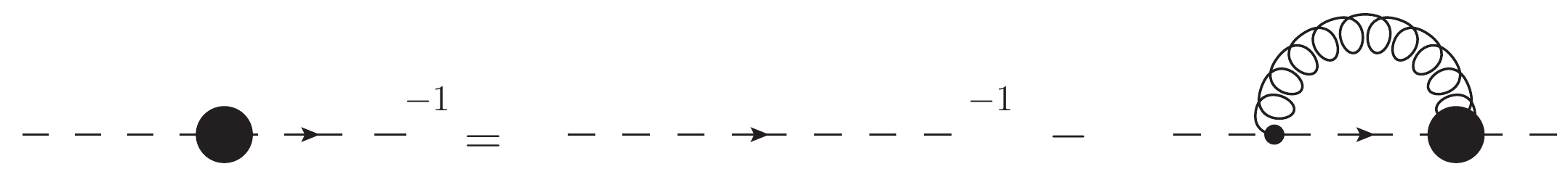}
\vskip5mm
\includegraphics[width=11cm]{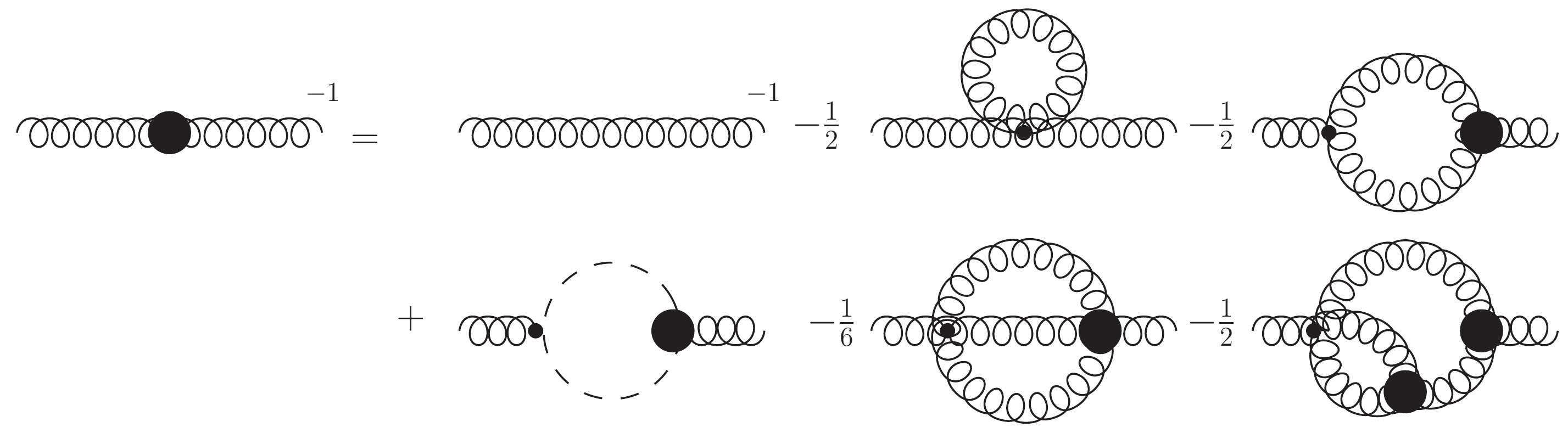}
\begin{center}
\caption{\label{fig:prop-DSEs}Full two-point Dyson-Schwinger equations of Landau gauge Yang-Mills theory. All internal propagators are dressed. Thick blobs denote dressed vertices. Wiggly lines are gluons, dashed ones ghosts.}
\end{center}
\end{figure}

The DSEs for the ghost and gluon dressing functions, $G(p^2)$ and $Z(p^2)$, respectively, read
\begin{align}
\label{eq:gh-DSE}
 \frac{1}{G(p^2)}&=Z_3+N_c\,g^2\,\tilde{Z}_1\,\int_q Z(q^2)G((p+q)^2) K_{G}(p,q)\Gamma^{A\bar{c}c}(q;p+q,p)\\
\label{eq:gl-DSE}
 \frac{1}{Z(p^2)}&=\tilde{Z}_3+N_c\,g^2\,\tilde{Z}_1\,\int_q G(q^2)G((p+q)^2) K_{Z}^{gh}(p,q)\Gamma^{A\bar{c}c}(p;p+q,q)\nnnl
 &+N_c\,g^2\,Z_1\,\int_q Z(q^2)Z((p+q)^2) K_{Z}^{gl}(p,q)\Gamma^{A^3}(p,q,-p-q),
\end{align}
where $\int_q$ stands for $\int d^4q/(2\pi)^4$ and $\tilde{Z}_1$ and $Z_1$ are the renormalization constants of the ghost-gluon and three-gluon vertices, respectively, whereas $\tilde{Z}_3$ is the renormalization constant of the ghost propagator and $Z_3$ that for the gluon propagator.
The quantities $\Gamma^{A\bar{c}c}(k;p,q)$ and $\Gamma^{A^3}(k,p,q)$ are dressing functions of the ghost-gluon and three-gluon vertices, respectively. The two DSEs are shown diagrammatically in \fref{fig:prop-DSEs}. The ghost and gluon propagators $D_{gh}(p^2)$ and $D_{gl,\mu\nu}(p^2)$ and their dressing functions are related by
\begin{align}
 D_{gh}(p^2):=-\frac{G(p^2)}{p^2}, &\quad  D_{gl,\mu\nu}(p^2):=P_{\mu\nu}(p)\frac{Z(p^2)}{p^2}.
\end{align}
The kernels $K_{G}$, $K_Z^{gh}$ and $K_Z^{gl}$ are given explicitly in Appendix~\ref{sec:app_kernels}. The coupled system for the two propagators was investigated many times in the literature with functional methods, see, for example, \cite{vonSmekal:1997is,vonSmekal:1997vx,Atkinson:1997tu,Zwanziger:2001kw,Lerche:2002ep,Zwanziger:2002ia,Fischer:2002hn,Pawlowski:2003hq,Zwanziger:2003cf,Aguilar:2008xm,Alkofer:2008jy,Fischer:2008uz,Fischer:2009tn,Huber:2009tx,Pennington:2011xs,LlanesEstrada:2012my}. Recently it has even been solved in the complex $p^2$-plane to extract the corresponding spectral functions for gluons and ghosts in complete agreement with local quantum field theory but with the expected positivity violations \cite{Strauss:2012dg}.  For complementary Euclidean results from Monte Carlo simulations see, e.g., \cite{Bloch:2003sk,Bogolubsky:2005wf,Sternbeck:2005tk,Ilgenfritz:2006he,Cucchieri:2006xi,Sternbeck:2006cg,Bogolubsky:2007bw,Cucchieri:2007rg,Cucchieri:2007md,Oliveira:2007dy,Bogolubsky:2007ud,Cucchieri:2008fc,Oliveira:2010xc,Oliveira:2012eh,Ayala:2012pb,Sternbeck:2012mf}.

In Section~\ref{sec:results_ghg} we will compare our results also to lattice data. Thus we shortly want to recall a few facts concerning the calculation of propagators on the lattice. It is often claimed that all lattice results show an unequivocal picture for the propagators. However, several open issues exist that need clarification. For example, it is well known that Gribov copies have an effect on the propagators \cite{Cucchieri:1997dx,Bogolubsky:2005wf,Sternbeck:2006rd,Maas:2009se,Maas:2011se}. An interpretation of this in terms of non-perturbative gauge completion was put forward in Ref.~\cite{Maas:2009se}. With functional equations one finds two types of solutions \cite{Boucaud:2008ji,Fischer:2008uz,Alkofer:2008jy} characterized by the value of the ghost dressing function at zero momentum: If it is finite, the family of decoupling solutions emerges which has a non-zero gluon propagator at zero momentum \cite{Dudal:2008sp,Boucaud:2008ji,Aguilar:2008xm,Fischer:2008uz,Alkofer:2008jy}. The second type is the scaling solution, for which the ghost dressing diverges and the gluon propagator vanishes \cite{vonSmekal:1997is}. Of course it is tempting to look for a direct correspondence between the choice of Gribov copies and the family of solutions. However, Monte Carlo methods sample different Gribov copies in a way not accessible to functional methods, so that unveiling such a relation is in no way trivial \cite{Maas:2013vd}. Attempts to deal with Gribov copies with continuum methods include the Gribov-Zwanziger framework \cite{Gribov:1977wm,Zwanziger:1989mf,Vandersickel:2012tz} and its "refined" \cite{Dudal:2008sp,Dudal:2007cw,Dudal:2011gd} or "alternative refined" form \cite{Gracey:2010cg} and that of Ref.~\cite{Serreau:2012cg} where an averaging over Gribov copies is performed. An important observation in this respect is that the form of functional equations is not changed by restriction to the first Gribov region \cite{Zwanziger:2003cf,Zwanziger:2001kw} as automatically done in Monte-Carlo simulations. Using the Gribov-Zwanziger action for functional equations confirmed that the IR part of the solution remains unaffected by such a restriction \cite{Huber:2010cq,Huber:2009tx}. Recently promising new results in the opposite direction, i.~e., selecting Gribov copies on the lattice such that results change as expected from functional methods, were presented \cite{Sternbeck:2012mf}: Choosing Gribov copies by the value of the lowest eigenvalue of the Faddeev-Popov operator seems to lead to changes in the propagators akin to those obtained when varying the boundary conditions in the DSEs. Another point in need of better understanding is the behavior in the strong coupling regime, for which all momenta can be considered below $\Lambda_{QCD}$ and thus in the IR. There two solution branches are found with characteristics of decoupling and scaling solutions \cite{Sternbeck:2008mv,Cucchieri:2009zt,Maas:2009ph}. Due to these ambiguities it is in principle important to consider the details of the gauge-fixing algorithm used to sample the Gribov copies of Landau gauge on the lattice when comparing the infrared behavior of the propagators with lattice data. All lattice data used for comparisons herein were obtained from sampling, with some algorithm-specific bias, local minima of the standard gauge-fixing potential for lattice Landau gauge thus yielding decoupling.

\subsection{The ghost-gluon vertex Dyson-Schwinger equations}
\label{sec:ghg}

\begin{figure}[tb]
 \includegraphics[width=\textwidth]{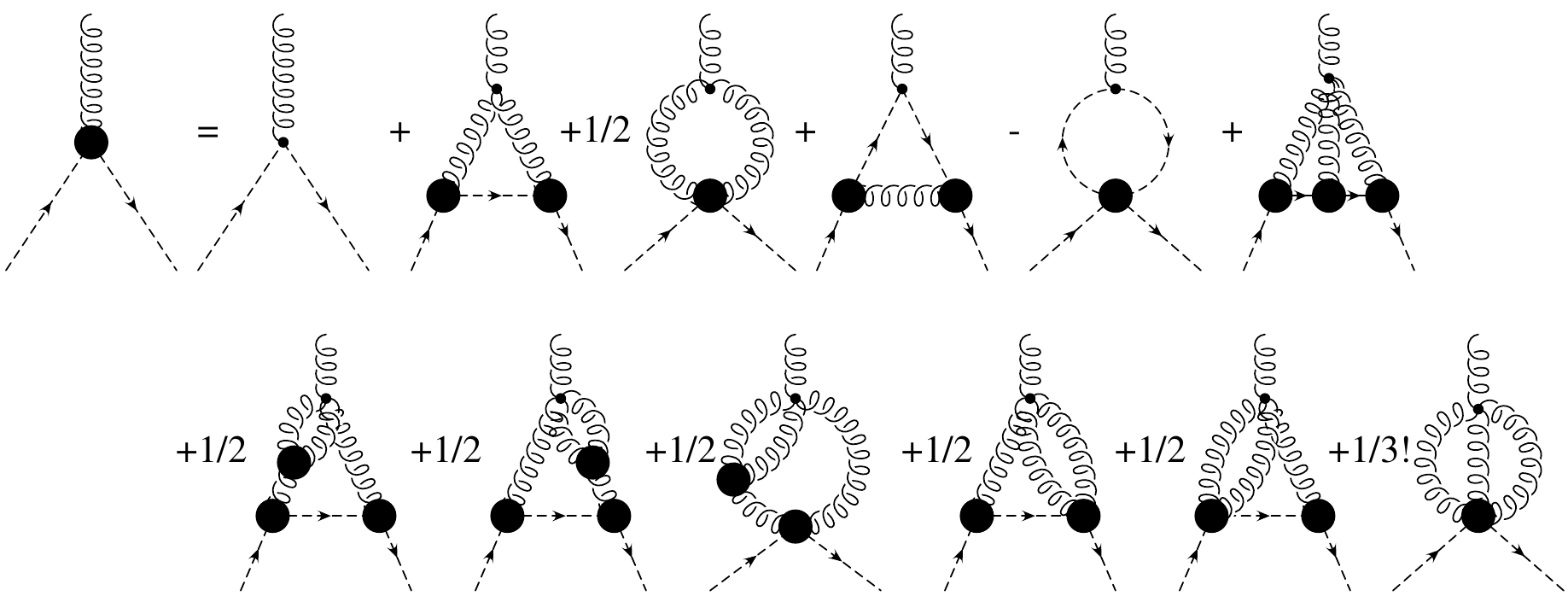}\\
 \includegraphics[width=0.7\textwidth]{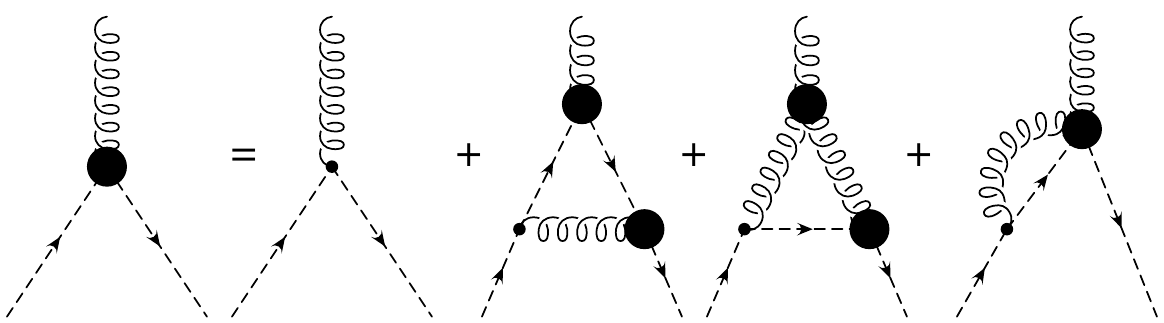}
\caption{\label{fig:ghg-DSE}The ghost-gluon vertex DSEs. All internal propagators are dressed. \textit{Top:} Version with the bare vertices attached to the gluon leg. \textit{Bottom:} Version with bare vertices attached to the ghost leg. Thick blobs denote dressed vertices. Wiggly lines are gluons, dotted ones ghosts.}
\end{figure}

The ghost-gluon vertex has two different DSEs which differ by the field that is attached to the bare vertex, see \fref{fig:ghg-DSE}. Of course both full equations are equivalent and we can choose the one most suitable for our calculations. It seems tempting to employ the DSE with the ghost leg attached to the bare vertex, because it has only four diagrams. However, both UV and IR consistent truncations depend on the dressed three-gluon vertex for which we use a model. To make our calculations as much model independent as possible we thus want to avoid the dressed three-gluon vertex in the ghost-gluon vertex DSE. We therefore consider the first DSE in \fref{fig:ghg-DSE}.

In the UV the leading diagrams are the two triangle diagrams. The two other one-loop diagrams, the gluon and the ghost loops with quartic ghost-gluon and ghost interactions, respectively, are UV subleading because the four-point functions therein are not primitively divergent. At the same time the triangle diagrams contain the leading IR corrections. This can be seen most easily for the scaling type solution: There, the two IR leading diagrams as indicated by the power counting scheme \cite{Alkofer:2004it} are the two diagrams with a bare ghost-gluon vertex. However, one of them contains a quartic ghost vertex which behaves as $(p^2)^\ka$ in the IR. We expect the contribution from this diagram to be suppressed as compared to the triangle diagram with a bare ghost-gluon vertex because there is no tree-level four-ghost coupling in Landau gauge. It corresponds to a two-loop contribution as can be seen explicitly by inserting the DSE of the quartic ghost vertex. For the decoupling type solution no IR enhancements of the vertices are expected \cite{Alkofer:2008jy}, so the main corrections should also come from the one-loop terms.
Note that this truncation is almost the same as the first three diagrams of the other ghost-gluon vertex DSE depicted at the bottom of \fref{fig:ghg-DSE} except for the dressed three-gluon vertex. Consequently all information of the diagrams neglected in our truncation is contained in the full three-gluon vertex and the ghost-gluon scattering kernel.

The dressed ghost-gluon vertex is described by two dressing functions. A typical choice is
\begin{align}
 \Gamma^{A\bar{c}c,abc}_\mu(k;p,q):=f^{abc}\Gamma^{A\bar{c}c}_\mu(k;p,q):=i\,g\,f^{abc}\left(p_\mu A(k;p,q)+k_\mu B(k;p,q)\right).
\end{align}
Here the momenta $k$, $p$, $q$ refer to the gluon, anti-ghost and ghost legs, respectively, as indicated in the superscript of $\Gamma^{A\bar{c}c}$. By convention we have all momenta ingoing. The dressing $A(k;p,q)$ is symmetric under exchange of the ghost momenta, $A(k,p,q)=A(k;q,p)$, but $B(k;p,q)$ is not \cite{Lerche:2002ep}. In the Landau gauge only $A(k;p,q)$ contributes to Green functions due to the transversality of the gluon propagator. However, it is slightly misleading to call $A(k;p,q)$ the transverse dressing of the ghost-gluon vertex, as is often done. A clear separation between transverse and longitudinal parts can be achieved with \cite{Lerche:2002ep,Huber:2012zj}:
\begin{align}
 \Gamma^{A\bar{c}c,abc}_\mu(k;p,q):=i\,g\,f^{abc}\left(P_{\mu\nu}(k)p_\nu D^{A\bar{c}c}_t(k;p,q)+k_\mu D^{A\bar{c}c}_l(k;p,q)\right),
\end{align}
where $P_{\mu\nu}(k)$ is the transverse projector $g_{\mu\nu}-k_\mu k_\nu/k^2$.
The relations between the two sets of dressings are
\begin{align}
 D^{A\bar{c}c}_t(k;p,q)&=A(k;p,q),\\
 D^{A\bar{c}c}_l(k;p,q)&=B(k;p,q)+A(k;p,q)\,p\cdot k/k^2
\end{align}
from which one sees that $A(k;p,q)$ also appears in the longitudinal dressing function and that the transverse part is symmetric under exchange of the ghost momenta, $D^{A\bar{c}c}_t(k;p,q)=D^{A\bar{c}c}_t(k;q,p)$.
The bare vertex is obtained for $A(k;p,q)=1$, $B(k;p,q)=0$ and $D^{A\bar{c}c}_t(k;p,q)=1$, $D^{A\bar{c}c}_l(k;p,q)=k\cdot p/k^2$, respectively. Note that the choice of a ghost-gluon vertex that ensures the IR transversality of the gluon DSE as in Refs. \cite{Lerche:2002ep,Fischer:2008uz} corresponds to setting the longitudinal dressing $D^{A\bar{c}c}_l(k;p,q)$ to zero in the IR. However, when using the transversely projected gluon DSE $D^{A\bar{c}c}_l(k;p,q)$ does not enter anyway and only appears when testing if the longitudinal components of the gluon propagator vanish.

Using the transverse/longitudinal basis one can easily see that the ghost-gluon vertex has no transverse part for zero ghost momentum purely for kinematical reasons:
\begin{align}\label{eq:ghg_ghMom0}
 \Gamma^{A\bar{c}c}_\mu(-p;p,0)=i\,g\left(P_{\mu\nu}(-p)p_\nu D^{A\bar{c}c}_t(-p;p,0)-p_\mu D^{A\bar{c}c}_l(-p;p,0)\right)=-i\,g\,p_\mu D^{A\bar{c}c}_l(-p;p,0).
\end{align}
Note that this makes no statement about the transverse dressing function. We come back to this below.
This property of the ghost-gluon vertex is useful for the so-called $\widetilde{MOM}_h$ scheme \cite{Chetyrkin:2000fd} which is an asymmetric  momentum subtraction scheme defined by subtracting the radiative corrections to the ghost-gluon vertex at vanishing incoming ghost momentum. Since the vertex is bare in this limit, its renormalization constant $\tilde{Z}_1^{\widetilde{MOM}_h}$ is $1$ in Landau gauge, which was explicitly confirmed up to three loops in \cite{Chetyrkin:2000dq}.  
In terms of $A$ and $B$ \eref{eq:ghg_ghMom0} reads
\begin{align}
 \Gamma^{A\bar{c}c}_\mu(-p;p,0)=i\,g\,f^{abc}p_\mu\left( A(-p;p,0)- B(-p;p,0)\right).
\end{align}
This observation is related to what has become known as Taylor's non-renormalization argument \cite{Taylor:1971ff}. However, the argumentation in Ref.~\cite{Taylor:1971ff} was different. Taylor started from the following parametrization of the vertex, which is based on the transversality of the gluon propagator:
\begin{align}
 \Gamma^{A\bar{c}c}_\mu(k,p,q)=i\,g\, p_\mu+p_\nu q_\rho F_{\mu\nu\rho}(k,p,q).
\end{align}
For $p^2=q^2=k^2=0$ the second term should equal zero because of the factor $q_\rho$ \cite{Marciano:1977su} leading to $\tilde{Z}_1=1$ if the vertex is renormalized by momentum subtraction at $p^2=q^2=k^2=0$.\footnote{Note that $\tilde{Z}_1=1$ is automatically true for renormalization schemes based on minimal subtraction like $\overline{MS}$ \cite{Boucaud:2005ce} or \textit{MiniMOM} \cite{vonSmekal:2009ae}, because all diagrams are finite. In the Landau gauge the $\widetilde{MOM}_h$ and the \textit{MiniMOM} schemes are thus equivalent  \cite{vonSmekal:2009ae}.} However, this statement implicitly assumes that the value of the ghost-gluon vertex is the same independent from which direction the IR is approached. For the scaling solution it was already observed in Refs.~\cite{Schleifenbaum:2004id,Schleifenbaum:2006bq,Alkofer:2008dt} that the limits $p^2,\,q^2,\,k^2 \rightarrow 0$ and $q^2 \rightarrow 0$, $p^2\rightarrow 0$ are not equal. Furthermore, any possible singularities in $F_{\mu\nu\rho}(k,q,p)$ are disregarded. So there seem to be some caveats in the original argument why the ghost-gluon vertex becomes bare in the limit of vanishing incoming ghost momentum $q$. A third way to obtain this result is based on the ghost-gluon vertex Slavnov-Taylor identity (STI) \cite{vonSmekal:1997vx}:
\begin{align}
\frac{1}{\xi}\partial_\mu^y \langle A_\mu^b(y)\bar{c}^c(z)c^a(x)\rangle - \frac{1}{\xi}\partial_\mu^z \langle A_\mu^c(z)\bar{c}^b(y)c^a(x)\rangle=\frac{1}{2}g\,f^{ade}\langle c^d(x)c^e(x)\bar{c}^c(z)\bar{c}^b(y)\rangle.
\end{align}
If one ignores the connected part of the ghost scattering kernel one arrives at
\begin{align}
 i\,k_\mu \Gamma_\mu(k;p,q)G^{-1}(k^2)+i\,q_\mu \Gamma_\mu(q;p,k)G^{-1}(q^2)=g\,p^2G^{-1}(p^2).
\end{align}
If the connected part is maintained, the following conclusions remain consistent but a rigorous proof is lacking.
Of course, since this is an STI, it only provides information about the longitudinal part of the ghost-gluon vertex in agreement with the observation in \eref{eq:ghg_ghMom0}. Assuming that the longitudinal dressing function $D_l^{A\bar{c}c}(k;p,q)$ has no strong divergence for $k\rightarrow 0$, we can take $q$ or $k$ to zero and use $\lim_{q^2\rightarrow 0} q^2 G^{-1}(q^2)=0$, which is valid for both decoupling and scaling solutions, to obtain
\begin{align}\label{eq:Dlq0}
 \,D^{A\bar{c}c}_l(-p;p,0)=-1.
\end{align}
This is equivalent to
\begin{align}
 \Gamma_\mu(-p;p,0)=i\,g \,p_\mu,
\end{align}
i.e., the ghost-gluon vertex reduces to the tree-level contribution in this limit as expected. From \eref{eq:Dlq0} we can switch to the dressings $A(k;p,q)$ and $B(k;p,q)$ and obtain
\begin{align}\label{eq:ABq0}
 B(-p;p,0)-A(-p;p,0)=-1.
\end{align}
This relation was noted also, for example, in Ref.~\cite{Boucaud:2005ce}.
Interestingly, it also means that the transverse dressing is not necessarily zero, because of $D^{A\bar{c}c}_t(-p;p,0)=A(-p;p,0)$. Thus when we calculate $A(k;p,q)$ or $D^{A\bar{c}c}_t(k;p,q)$ it may \textit{not} vanish for $q\rightarrow 0$ and the vanishing of the radiative corrections to the ghost-gluon vertex for zero ghost momentum is only of kinematic origin.

In order to calculate the ghost-gluon vertex one can either project out the dressings to obtain scalar equations, which is most easily done for the $D^{A\bar{c}c}_{t/l}$ basis, or decompose the appearing vector integrals into scalar ones, which is the method of choice for  the $A/B$ basis. Here we follow the latter approach.
The truncated DSE for the vertex reads then
\begin{align}
\label{eq:ghg-DSE}
 A&(k;p,q)=\nnnl
 &1+\frac{N_c\,g^2}{\Delta(p,k)}\int_r K^{A\bar{c}c}_{1}(k,p,r) D_{gh}(r^2)D_{gh}((r+k)^2)D_{gl}((p-r)^2)A(r-p; p, -r) A(p-r; k+r, -k-p)\nnnl
 &+\frac{N_c\,g^2}{\Delta(p,k)} \int_r K^{A\bar{c}c}_{2}(k,p,r) D_{gh}((r-p)^2)D_{gl}(r^2)D_{gl}((r+k)^2)A(r;p,p-r)A(k+r;p-r,-k-p)
\end{align}
with $\Delta(p,k)=p^2 k^2-(p\cdot k)^2$.
The kernels $ K^{A\bar{c}c}_{1}(k,p,r)$ and $ K^{A\bar{c}c}_{2}(k,p,r)$ are given in Appendix~\ref{sec:app_kernels}. Since the dressing $B(k;p,q)$ is not required for the Landau gauge, we do not give the corresponding expressions.

\subsection{Three-gluon vertex}

For the three-gluon vertex we will employ an ansatz. In the past different expressions have been used. The main guideline in their construction was the correct UV behavior for the gluon propagator, but also other aspects like quadratic UV divergences in the gluon DSE played a role, see Sec. \ref{sec:quad-divs} and, for instance, Refs.~\cite{vonSmekal:1997vx,Fischer:2003zc,Fischer:2002hn,Fischer:2002eq,Fischer:2008uz}. Since we use the transversely projected gluon two-point DSE we only deal with the completely transverse part of the vertex, which is unrestricted by Slavnov-Taylor identities. Quantitatively the vertex is expected to contribute to the closing of the gap between DSE and lattice results for the gluon dressing function in the mid-momentum regime. Another source of missing strength there are two-loop diagrams, of which one, the so-called squint diagram (the last one in \fref{fig:prop-DSEs}) also contains the three-gluon vertex. Using the functional renormalization group the propagators can be calculated such that this gap practically vanishes \cite{Fischer:2008uz}, but due to the different structure of the corresponding equations, it can not be inferred how much additional support in the mid-momentum regime is expected from which source in DSEs: two-loop diagrams or the dressing of the three-gluon vertex in the gluon loop. First calculations of two-loop diagrams in DSEs were performed in \cite{Bloch:2003yu,Alkofer:2011di}.

The IR behavior of the three-gluon vertex shows some interesting features. For the scaling type solution it was found that the uniform IR limit, i.e., when all momenta approach zero simultaneously, obeys the power law $(p^2)^{-3\ka}$ \cite{Alkofer:2004it} and that possible kinematic soft divergences can occur that go like $1-2\ka$ \cite{Alkofer:2008jy}. However, for the transversely projected vertex the degree of divergence is reduced to $3/2-2\ka$ \cite{Alkofer:2008dt,Fischer:2009tn} and such divergences play no role. For decoupling type solutions no IR divergences in the three-gluon vertex are expected \cite{Alkofer:2008jy}.

In light of the discussion of possible ambiguities in the IR behavior of the solutions of Landau gauge Yang-Mills theory it is interesting to consider the three-gluon vertex as obtained with Monte Carlo simulations: While in four dimensions the data does not show a zero crossing in the IR \cite{Alles:1996ka,Boucaud:1998bq,Cucchieri:2008qm}, probably because of too small lattices, the data in three dimensions becomes negative at low momenta \cite{Cucchieri:2008qm}. Also in two dimensions \cite{Maas:2007uv} a zero crossing is found and additionally the data agrees with the expected power law from a scaling solution \cite{Huber:2007kc, Huber:2012zj}. This is not unexpected, because it has become by now clear that in two dimensions no decoupling solution exists \cite{Cucchieri:2012cb,Huber:2012zj,Zwanziger:2012xg}. However, in three and four dimensions only the decoupling type solution is found on the lattice and hence the realization of such a power law would indeed be unexpected. Since no IR singularities are expected \cite{Alkofer:2008jy}, the vertex most likely approaches a constant but negative value at zero momenta. Further lattice results are therefore desirable for clarification of the exact IR behavior of the vertex. Based on the qualitative similarity of the propagators of three and four dimensions we assume that the zero crossing also exists in four dimensions and it is therefore included in our vertex model here as well. Further evidence in support of this is presented in Sec.~\ref{sec:results_3g} from a DSE calculation of the three-gluon vertex.

The employed construction of the three-gluon vertex is based on the one introduced in Refs. \cite{Fischer:2003zc,Fischer:2002hn,Fischer:2002eq}:
\begin{align}\label{eq:3g-Fischer}
 D^{A^3}(p,q,-p-q)=\frac{1}{Z_1}\frac{\left[G(y)G(z)\right]^{1-a/\delta-2a}}{\left[Z(y)Z(z)\right]^{1+a}},
\end{align}
where $Z_1$ is the renormalization constant of the three-gluon vertex, $y=q^2$, $z=(p+q)^2$ and $\de=-9/44$ is the anomalous ghost dimension, related to that of the gluon by $1+\gamma+2\de=0$. $a$ is a parameter which is irrelevant in the UV where it drops out. A typical choice is $a=3\delta$, which renders the dressing finite at vanishing momenta in case of the scaling solution \cite{Fischer:2003zc}. For the decoupling type solution the corresponding choice is $a=-1$. For our construction described below two observations are important: First, this vertex construction is not Bose symmetric; the momenta $p$ and $q$ are the external and internal momenta, respectively, in the gluon loop. Secondly, it contains the inverse of the renormalization constant of the three-gluon vertex to cancel the $Z_1$ appearing in the gluon loop of the gluon DSE due to the bare three-gluon vertex. By construction this model ensures the corrected logarithmic running of the gluon loop according to resummed perturbation theory. However, it also means that the anomalous dimension of the three-gluon vertex model itself is off by a factor of two. In earlier works this was remedied by replacing $Z_1$ by a momentum dependent function \cite{vonSmekal:1997vx}. Because the correct anomalous dimension requires also contributions of order $g^4$ and higher, this cannot be reproduced directly with the employed truncation of the gluon DSE and we will use a similar renormalization group improvement term.
Other models with Bose symmetry and the correct anomalous dimension of the vertex were used in Ref.~\cite{Alkofer:2008tt}, in the context of the quark-gluon vertex DSE, and in Ref.~\cite{Pennington:2011xs} for the Yang-Mills propagator system. However, in the latter the anomalous dimension of the gluon propagator was only reproduced by interpolating between these models in the IR and the one of \eref{eq:3g-Fischer} with $a=0$ in the UV \cite{Bloch:2001wz} so that in effect the employed model did not have the correct anomalous dimension.

We first address the issue of Bose symmetry.
Naively adding factors of $G(x)$ and $Z(x)$ with $x=p^2$ does not work, but instead we use a Bose symmetric combination of momenta:\footnote{We acknowledge discussions with Christian S. Fischer and Stefan Strauss who employ a similar construction.}
\begin{align}\label{eq:3g-UV}
D^{A^3,UV}(p,q,-p-q)=G\left(\frac{x+y+z}{2}\right)^{\alpha}Z\left(\frac{x+y+z}{2}\right)^{\beta}.
\end{align}
In order to determine $\alpha$ and $\beta$ we demand that the vertex has the correct anomalous dimension:
\begin{align}
 \delta\, \alpha+ \gamma\, \beta=\gamma_{3g}.
\end{align}
From the Slavnov-Taylor identity $Z_1=Z_3/\tilde{Z}_3$ we know $\gamma_{3g}=1+3\delta=17/44$. Furthermore we require IR finiteness, i.e.,
\begin{align}
 \de_{c}\,\alpha+\de_{A}\,\beta=0,
\end{align}
where $\de_{c}$ and $\de_A$ are the IR exponents of the ghost and gluon propagators, respectively.
This leads for the scaling solution to $\alpha=-2-6\de$ and $\beta=-1-3\de$. For decoupling we have $\alpha=3+1/\de$ and $\beta=0$.
Note that IR finiteness is not per se required, since we will control the IR behavior by another term, but for the current model it is advantageous that $D^{A^3,UV}(p,q,-p-q)$ does not vanish at zero momentum.

Next, we modify the IR behavior in order to incorporate the zero crossing expected from lattice calculations by the following expression:
\begin{align}\label{eq:3g_IR}
 D^{A^3,IR}(p,q,-p-q)=h_{IR} \,G(x+y+z)^{3}(f^{3g}(x)f^{3g}(y)f^{3g}(z))^4,
\end{align}
with the damping factors
\begin{align}
 f^{3g}(x):=\frac{\Lambda^2_{3g}}{\Lambda_{3g}^2+x}.
\end{align}
A similar construction to obtain the zero crossing was employed in two dimensions \cite{Huber:2012zj}. The exponent of the damping functions is $4$ in order to reproduce the fast change seen in three-dimensional lattice calculations. The impact of this IR part is controlled by the two parameters $h_{IR}$ and $\Lambda_{3g}$. Naturally $h_{IR}$ is negative to produce a zero crossing. From currently available lattice data \cite{Cucchieri:2008qm} one can estimate bounds for the parameters, which, however, is not a straightforward procedure because the conversion from internal to physical units can only be done a posteriori.
The total vertex employed in this study is given by
\begin{align}\label{eq:3g-new}
 D^{A^3}(p,q,-p-q)=D^{A^3,IR}(p,q,-p-q)+D^{A^3,UV}(p,q,-p-q).
\end{align}

Finally, we want to incorporate the correct UV behavior of the gluon propagator by adding a corresponding improvement term in the gluon loop. As mentioned above, this is necessary because the bare vertex in the gluon loop does not give any contribution to the anomalous dimension. Note that this problem would be absent, if we had both three-gluon vertices dressed. This is the case for the functional renormalization group \cite{Fischer:2008uz} and also for equations of motion obtained from nPI effective actions \cite{Berges:2004pu}. Earlier studies accounted for this in the construction of the three-gluon vertex model by using an anomalous dimension a factor two too large \cite{Fischer:2003zc,Fischer:2002hn,Fischer:2002eq,Fischer:2008uz}. However, since we want to use lattice calculations as input to improve on the vertex model, we have to split the renormalization group improvement term and the three-gluon vertex model. 
In analogy to the three-gluon vertex model we choose a Bose symmetric expression:
\begin{align}
 D^{A^3}_{RG}(p,q,-p-q)=\frac{1}{Z_1}D^{A^3,UV}(p,q,-p-q).
\end{align}
This term and the three-gluon vertex model are combined in the function $\Gamma^{A^3}(p,q,-p-q)$ in the gluon DSE:
\begin{align}
 \Gamma^{A^3}(p,q,-p-q)=D^{A^3}(p,q,-p-q) \,D^{A^3}_{RG}(p,q,-p-q)
\end{align}
Here it becomes obvious why we required IR finiteness of the UV part, because otherwise the whole expression could vanish for zero momenta.
The presented model for the three-gluon vertex is compared with lattice data in Section~\ref{sec:results_3g}.

\section{Methods}
\label{sec:methods}

\subsection{Quadratic divergences in the gluon DSE}
\label{sec:quad-divs}

The use of a UV regularization which is not gauge invariant entails that the gluon DSE is plagued by quadratic divergences. Several ways of dealing with them have been employed in the literature. For example, subtractions at the level of the integrands \cite{Fischer:2002eq,Fischer:2003zc} or modifications of the vertices \cite{Fischer:2008uz} were used. Another possibility is the projection with the so-called Brown-Pennington projector \cite{Brown:1988bn,Bloch:2001wz,Pennington:2011xs}, which, however, leads to the dependence on the longitudinal part of the ghost-gluon vertex. It also introduces  a spurious dependence on an additional gauge parameter \cite{Lerche:2002ep,Fischer:2010is}, which should be irrelevant in the Landau gauge. Thus we employ the usual transverse projection.

Here we use the following method to handle the quadratic divergences: Instead of adding an additional term in the kernel of the gluon loop to subtract the quadratic divergences of both loops as done, for instance, in Refs. \cite{Fischer:2002eq,Fischer:2003zc}, we add such terms for each loop separately. This has, for example, the advantage that truncations that take only the ghost loop into account can be used. Another benefit is that a standard fixed-point iteration becomes a viable solution method, because if the quadratic divergences are subtracted only via the gluon loop in this case, instabilities are introduced in the iterative procedure and no solution is found \cite{Huber:2011xc}. This made the Newton method a crucial aspect of most previous solution processes.

Since quadratic divergences are a problem of the UV regime, we want to avoid any interference with the IR part. Thus the additional terms are suppressed in the IR by a damping factor. The corresponding expressions can be found in Appendix~\ref{sec:props_UV}.

\subsection{Solution method}

The system of the three DSEs is split into two subsystems, which are iterated in turn until a given precision is reached. The first subsystem consists of the propagator DSEs, for which we employ the Newton method. This method is used throughout the literature and we refer for details, for example, to Refs.~\cite{Atkinson:1997tu,Maas:2005xh,Fischer:2002hn,Fischer:2003zc,Huber:2011xc}. Alternatively we used a direct fixed point iteration for all three equations. We solved the equations on CPUs, but note that recently it was demonstrated that implementing such a system on GPUs can lead to a significant speedup \cite{Hopfer:2012ht}.

For the ghost-gluon vertex a standard fixed point iteration is employed. Therefor we put the ghost-gluon dressing function $A(k;q,p)$ on a grid and use linear interpolation for intermediate points. If points outside the grid are encountered we take the value at the boundary, since the vertex is constant in the IR and the UV. The validity of this procedure can also be confirmed a posteriori from the results. As variables we choose the momenta squared of the gluon and the anti-ghost, $k^2$ and $q^2$, respectively, and the cosine of the angle, $c_t=\cos \theta$, between these two momenta. The allowed intervals for the variables are then $[0,\infty]$, $[0,\infty]$ and $[-1,1]$. If we used a basis with three momenta squared, the intervals would depend on the values of the other variables through momentum conservation which would complicate the calculations unnecessarily; hence this choice is here more advantageous. Since the dressing function of the ghost-gluon vertex has to be calculated on a three-dimensional grid, the computing time increases considerably when compared with the propagator system. Furthermore, the kernels are more complicated and the integrations are three- instead of two-dimensional, with two angle variables $\theta_1$ and $\theta_2$. For the propagators in total three integrations have to be performed, because the radial momentum integration is split at the value of the external momentum. The integration of the vertex requires similar splittings~\cite{Schleifenbaum:2004dt} at $r^2=p^2$ with $\theta_2=0$ and at $r^2=k^2$ with $\theta_1=\pi$ and $\theta_2=\pi-\theta$, where $r$ is the radial integration variable and $\theta_1$ and $\theta_2$ are the angle integration variables. Thus in total six integrations have to be done.

The two subsystems are iterated in turn until the required precision is reached.
The framework \textit{CrasyDSE} \cite{Huber:2011xc} was employed for all calculations. Color factors were used throughout for the gauge group $SU(3)$.

\section{Effects of the three-gluon vertex}
\label{sec:results_3g}

\begin{figure}[tb]
 \begin{center}
  \includegraphics[width=0.49\textwidth]{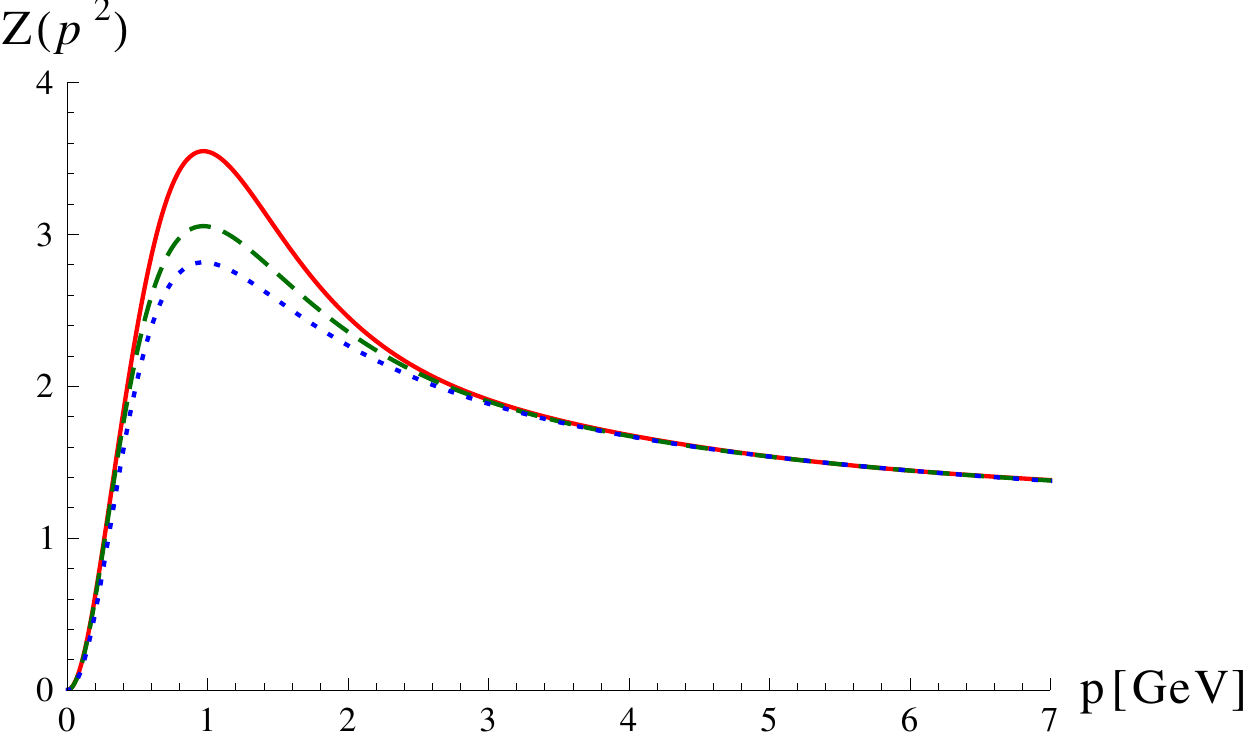}
  \includegraphics[width=0.49\textwidth]{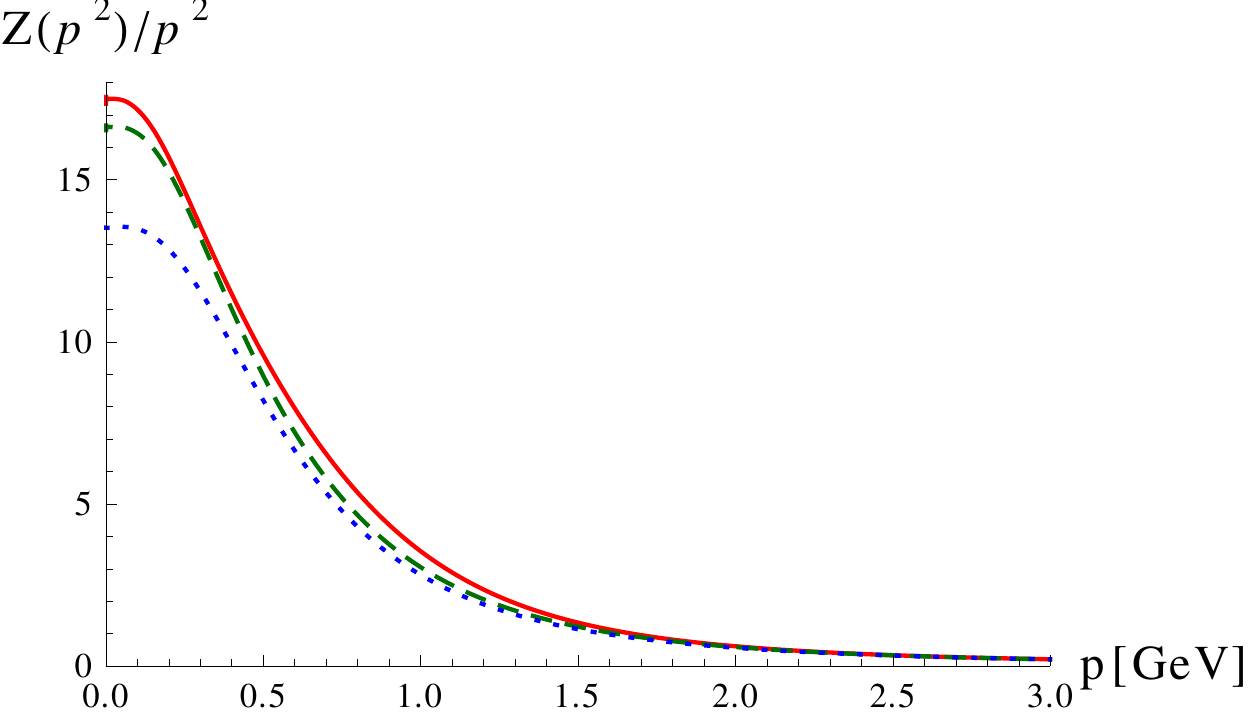}\\
  \includegraphics[width=0.49\textwidth]{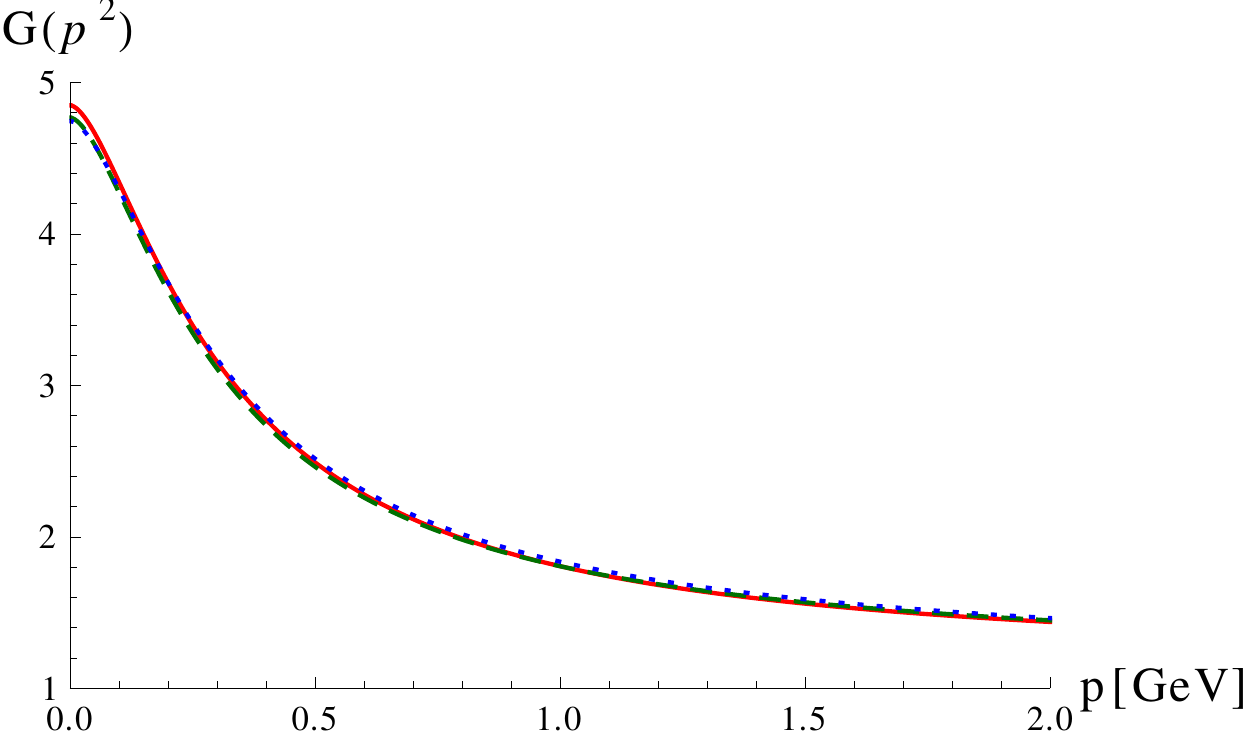}\hspace{0.49\textwidth}
  \caption{\label{fig:comp3gModels}Comparison of different three-gluon vertex models: model of \eref{eq:3g-Fischer} (blue/dotted line), Bose symmetrized model (\eref{eq:3g-new} with $h_{IR}=0$; green/dashed line), Bose symmetrized model with IR enhancement (\eref{eq:3g-new} with $h_{IR}=-1$, $\Lambda_{3g}=1.7\,GeV$; red/continuous line). \textit{Top:} Gluon dressing function (\textit{left}) and propagator (\textit{right}). \textit{Bottom:} Ghost dressing function, for which the three lines almost lie on top of each other.}
 \end{center}
\end{figure}

In this section we illustrate the quantitative effects of the three-gluon vertex model and investigate the impact of the position of its zero crossing by considering the system of propagators. The ghost-gluon vertex is set to tree-level. By construction the UV behavior of the propagators is not subject to any change. The most crucial modifications will be seen in the mid-momentum regime. For the IR only quantitative changes are expected, like a varying value of the gluon propagator at zero momentum.

We start by comparing three different three-gluon vertex models: The one of \eref{eq:3g-Fischer}, the Bose-symmetrized one (\ref{eq:3g-UV}), and the IR enhanced one, viz. \eref{eq:3g-new} with a non-zero $h_{IR}$. The results for a decoupling type solution are shown in \fref{fig:comp3gModels}. As can clearly be seen, both the Bose symmetrization and the IR enhancement enlarge the bump in the gluon dressing function. Also the zero value of the gluon propagator changes and raises to higher values. The ghost propagator, on the other hand, remains nearly unaffected. In Section~\ref{sec:results_ghg} we will compare the propagators also to lattice results and will see that the enlarged bump considerably reduces the gap between DSE and lattice results.

\begin{figure}[tb]
 \begin{center}
  \includegraphics[width=0.49\textwidth]{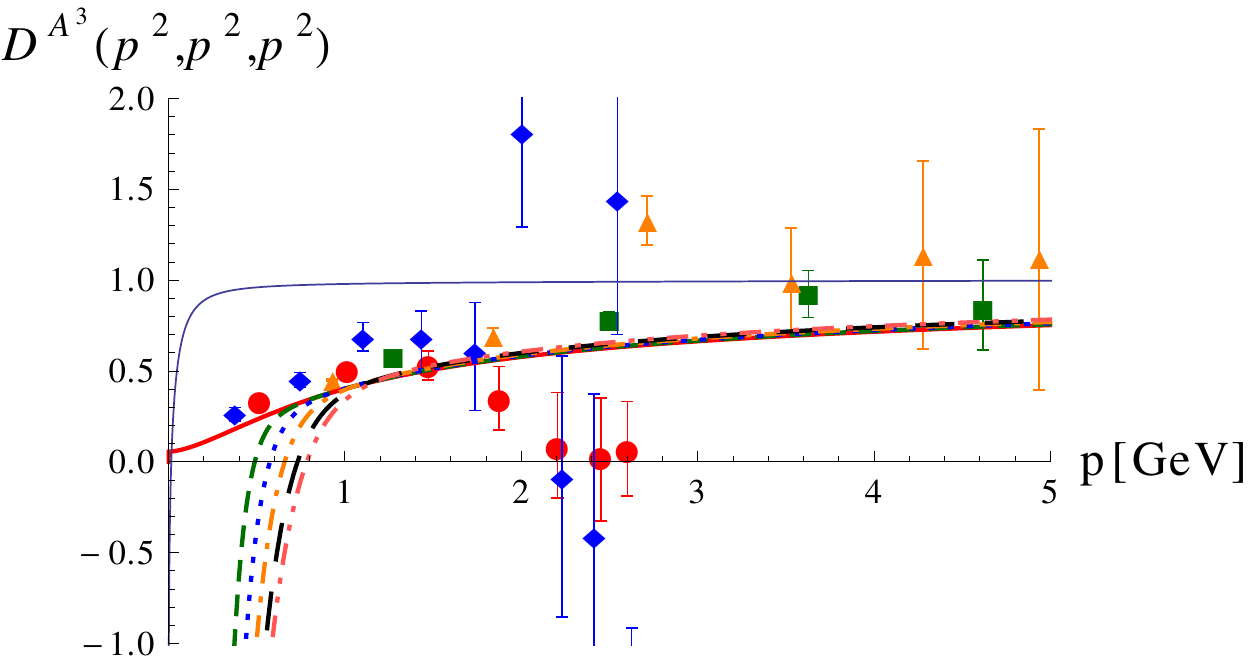}
  \includegraphics[width=0.49\textwidth]{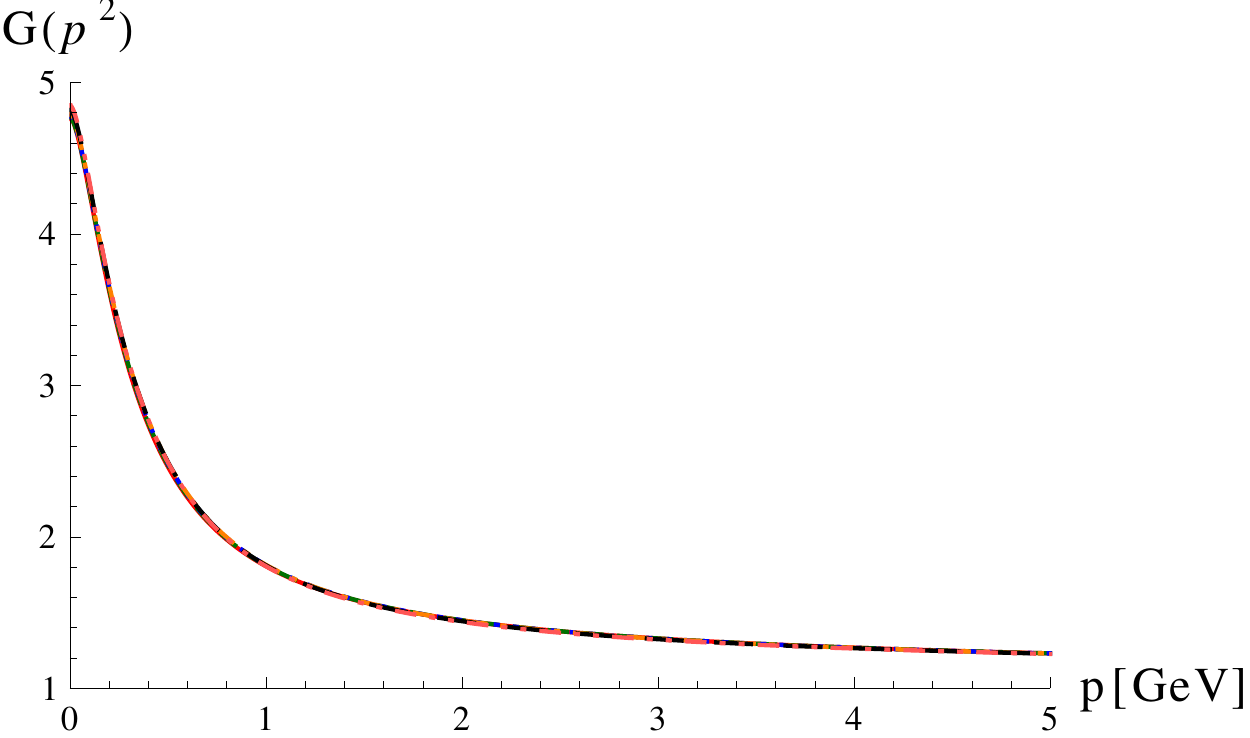}\\
  \includegraphics[width=0.49\textwidth]{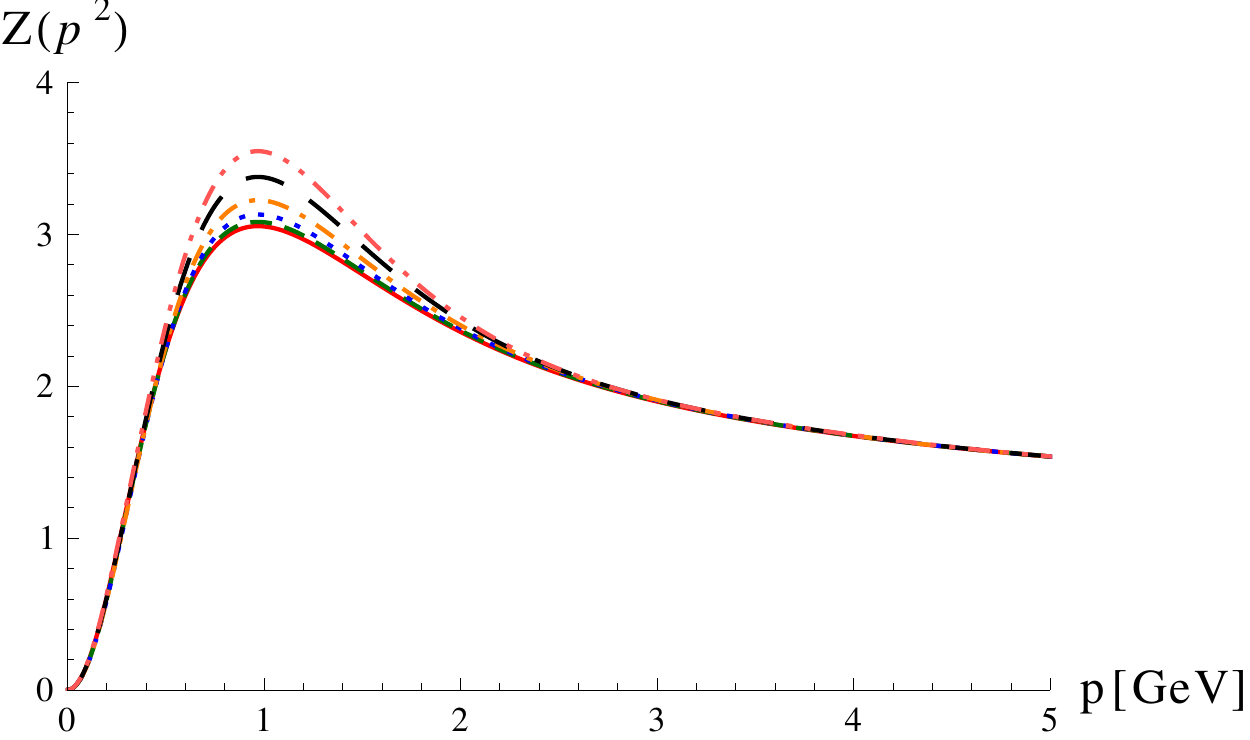}
  \includegraphics[width=0.49\textwidth]{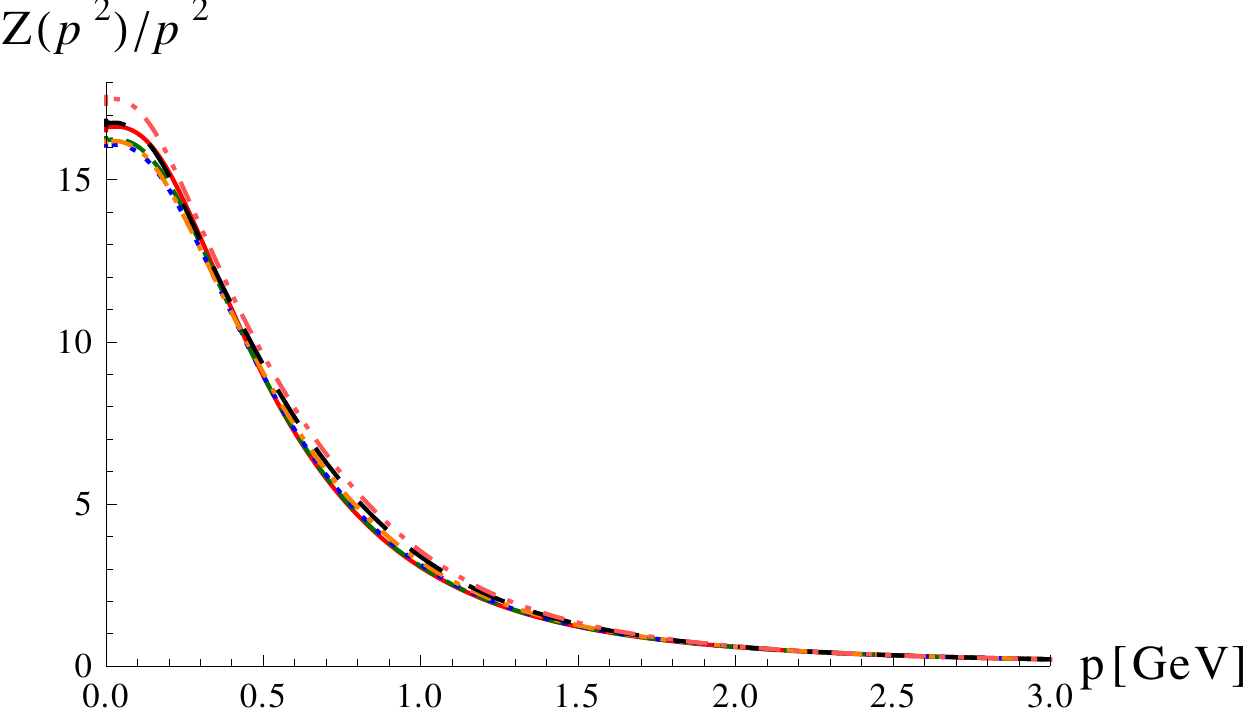}
  \caption{\label{fig:comp3gModel}Comparison of different parameters for the three-gluon vertex model: $h_{IR}=-1$, $\Lambda_{3g}=0,\,0.86,\,1.08,\,1.31,\,1.53,\,1.7\,GeV$.
\textit{Top left:} Symmetric configuration of the vertex model. Thick lines from left to right correspond to rising values of the parameter $\Lambda_{3g}$. The thin line is the IR leading contribution of the three-gluon vertex calculated from its DSE showing a zero crossing at very low momenta. Individual points are lattice data from Ref.~\cite{Cucchieri:2008qm}: Red circles correspond to $N=16$/$\beta=2.2$, green  squares to $N=16$/$\beta=2.5$, blue diamonds to $N=22$/$\beta=2.2$ and orange triangles to $N=22$/$\beta=2.5$.
\textit{Top right:} Ghost dressing function. All curves lie almost on top of each other.
\textit{Bottom:} Gluon dressing function (\textit{left}) and propagator (\textit{right}). Curves from low to high correspond to values of $\Lambda_{3g}$ from low to high.}
 \end{center}
\end{figure}

What we need for the three-gluon vertex model are values for its parameters. Ideally we could get them from lattice data. However, due the presently large uncertainties this is still somewhat ambiguous. Thus we test several values and their influence on the propagators. For simplicity we fix the parameter $h_{IR}=-1$ and only vary $\Lambda_{3g}$. In \fref{fig:comp3gModel} we show results for values of  $\Lambda_{3g}$ between $0$ and $1.7\,GeV$. The choice $\Lambda_{3g}=0\,GeV$ means that the IR enhanced part is dropped. In addition to the propagator dressing functions we also show a comparison of the three-gluon vertex model with lattice data at the symmetric point. This data is for $SU(2)$, but since it is known that the propagators look extremely similar for $SU(2)$ and $SU(3)$ \cite{Sternbeck:2007ug} one might expect the same for the three-gluon vertex. The two other available configurations (orthogonal momenta of equal magnitude, or one momentum vanishing) look similar. One can see from the lattice data that the turnover to negative values has to proceed rather rapidly. This is the reason why the exponent of the damping functions in \eref{eq:3g_IR} is chosen as $4$. For the plot of the three-gluon vertex in \fref{fig:comp3gModel} the final propagators were used in the model. As can be seen, the shown models have a zero crossing at higher values than indicated by the data. Such high values were chosen to demonstrate the effect it has on the gluon dressing function: Only for a zero crossing at momenta high enough the gluon dressing is affected. This is evident in the plot of the gluon dressing functions where the first three curves, corresponding to no zero crossing and crossings below  $600\,MeV$, show not much difference. For higher values of the zero crossing the bump increases faster. Again we see no effect on the ghost dressing function and only a small one on the zero momentum value of the gluon propagator.

The observations on the dependence on the place of the zero crossing lead to the conclusion that the actual value of the three-gluon vertex in the deep IR is not very important for the propagators. Indeed, for values of up to roughly $0.7\,GeV$ for $\Lambda_{3g}$ we do not find any difference to the propagators obtained with a three-gluon vertex without zero crossing. This is not too surprising considering that the three-gluon vertex in the gluon loop is multiplied by two gluon propagators. For the scaling solution this leads to an IR suppression of $(p^2)^\ka$ and for decoupling solutions to $(p^2)^2$.

In order to shed more light on the question of the existence of a zero crossing, we calculated the IR leading part of the three-gluon vertex for the symmetric momentum configuration. In the IR the so-called ghost triangle is the dominant contribution. Since it contains no three-gluon vertices there are no back coupling effects and the calculation consists of only one iteration. Of course, the mid-momentum and UV behavior are not described accurately in this rough truncation. Using the ghost propagators from the propagator calculations and bare ghost-gluon vertices, the ghost triangle indeed yields a negative contribution leading to a zero crossing. The inclusion of the other diagrams, which are suppressed in the IR, may change its position slightly but they will not make it disappear. A more detailed investigation of the three-gluon vertex was done in Ref.~\cite{Huber:2012zj} for two dimensions. There it was found that gluonic diagrams indeed do not affect the IR part. Note that the ghost propagator is hardly affected by the different three-gluon vertex models and consequently all curves for the ghost triangle contribution are essentially the same.

\section{Influence of a dynamical ghost-gluon vertex}
\label{sec:results_ghg}

The ghost-gluon vertex is central for any truncation scheme of functional equations in the Landau gauge, where usually a bare vertex is employed. Deviations from tree-level might yield quantitative corrections in the mid-momentum regime, because here the difference to a bare vertex is largest. In this section we add the ghost-gluon vertex dynamically to the set of equations to be solved in order to assess how much quantitative change this entails for the propagators.
For the three-gluon vertex we will use the model of \eref{eq:3g-new}.

\begin{figure}[tb]
 \begin{center}
  \includegraphics[width=0.49\textwidth]{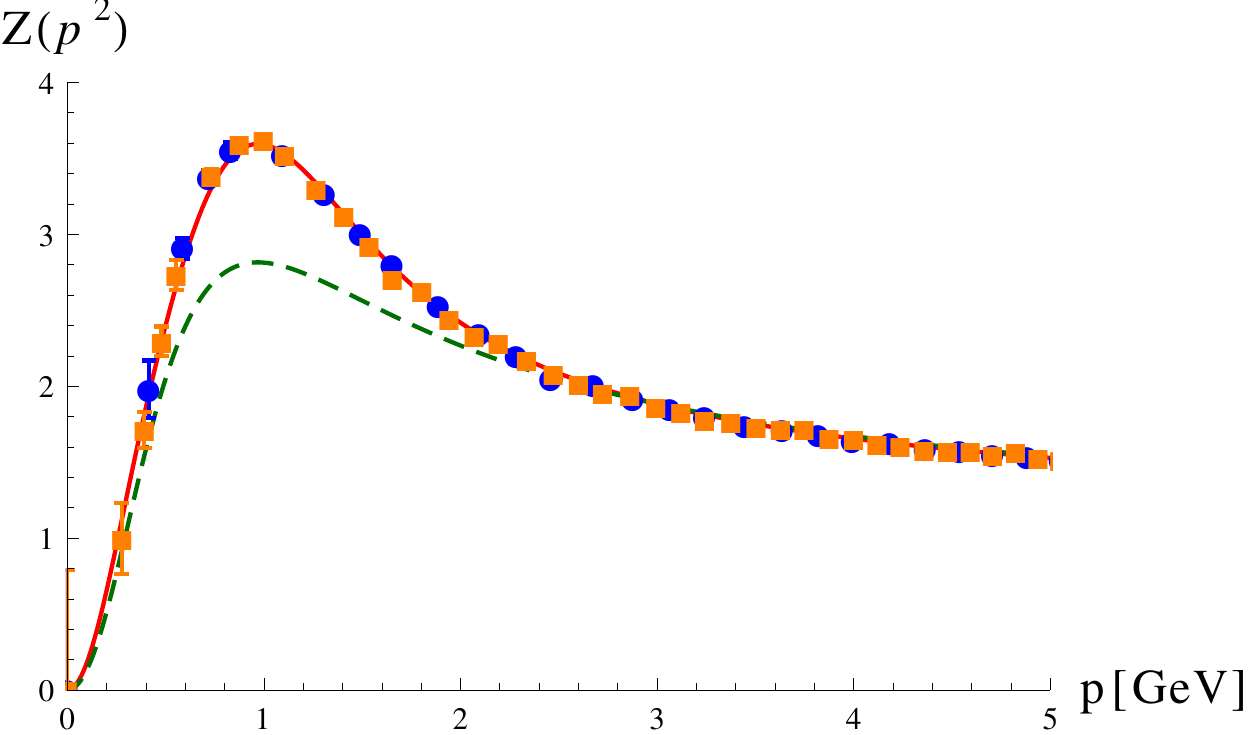}
  \includegraphics[width=0.49\textwidth]{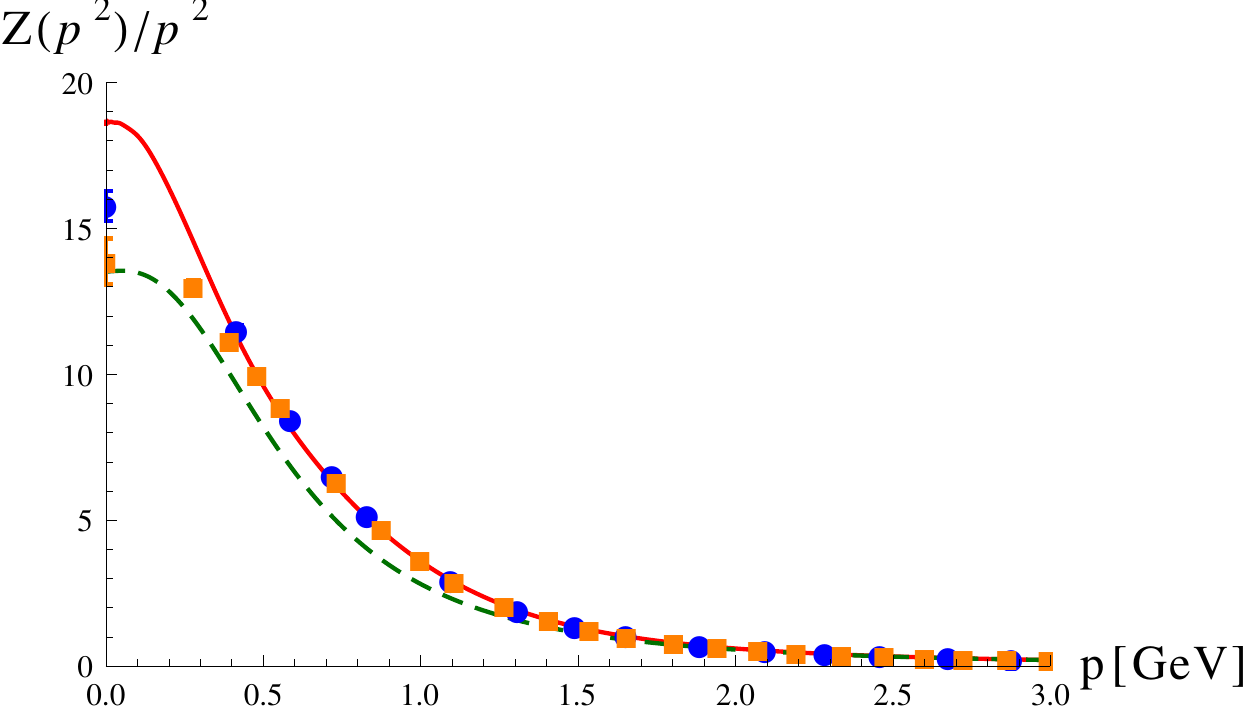}\\
  \includegraphics[width=0.49\textwidth]{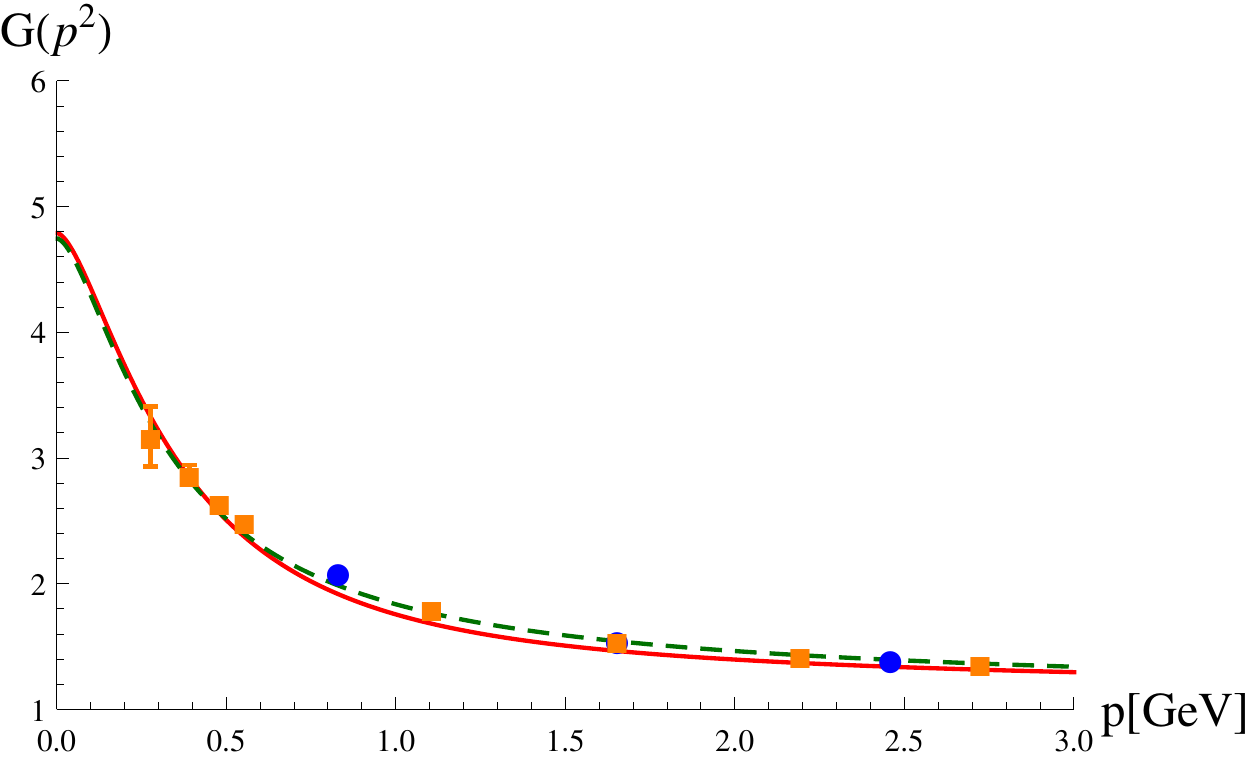}
  \includegraphics[width=0.49\textwidth]{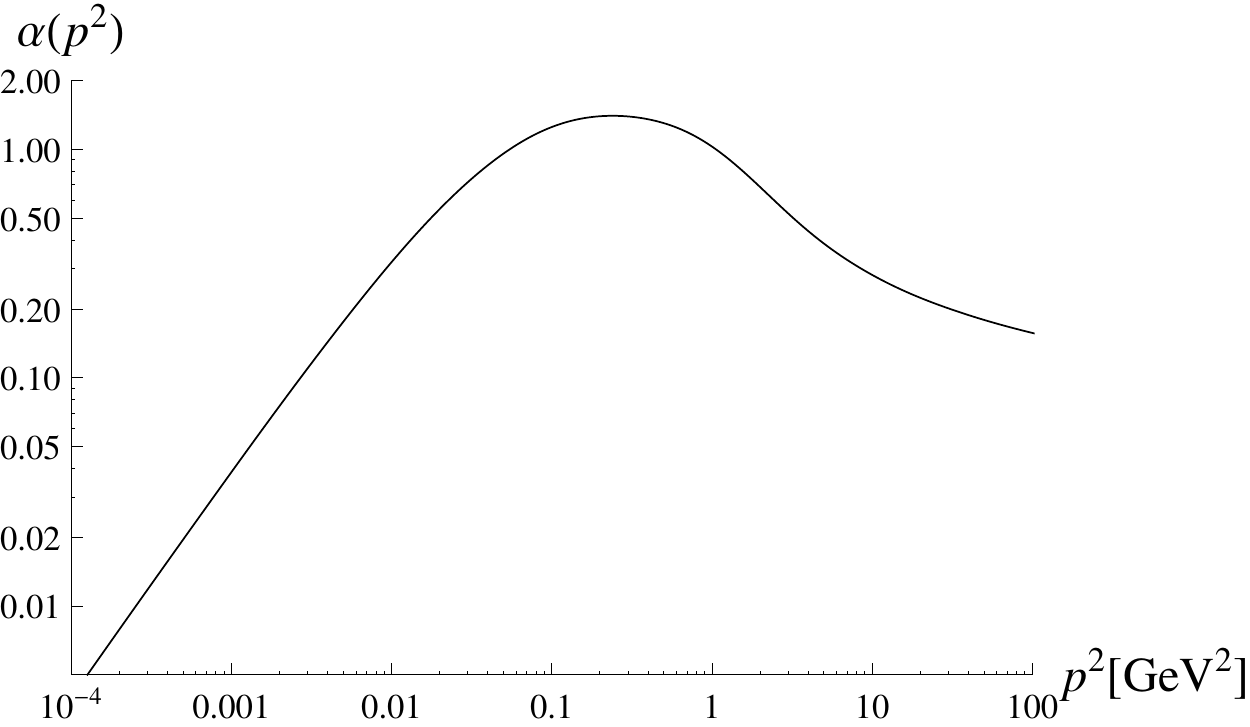}
  \caption{\label{fig:compLDynGhg_props}Comparison of the results for the gluon propagator dressing function $Z(p^2)$, the gluon propagator $Z(p^2)/p^2$ and the ghost dressing function $G(p^2)$ with lattice data \cite{Sternbeck:2006rd}. The red/continuous line was obtained with a dynamic ghost-gluon vertex and the optimized effective three-gluon vertex, \eref{eq:3g-new} with parameter set 2 of Tab.~\ref{tab:paras}. For reference the green/dashed line is shown, which was obtained with the three-gluon vertex model of \eref{eq:3g-Fischer} and a bare ghost-gluon vertex. Lattice data is for $\beta=6$ and lattice sizes of $L=32$ (blue circles) and $L=48$ (orange squares). The lower right plot shows the corresponding coupling calculated via \eref{eq:coupling} from the propagators obtained from the DSEs.}
 \end{center}
\end{figure}

\subsection{Propagators}
\label{sec:results_ghg_props}

In order to obtain a solution for the DSEs consisting of the propagator and the ghost-gluon vertex equations, the three-gluon vertex model plays a crucial role. For example, using the model of \eref{eq:3g-Fischer} it turns out that the gluon loop can dominate over the ghost loop in the gluon DSE which results in a negative gluon dressing function. Thus the changes induced by the Bose symmetrization and the IR part are required in order to obtain a solution. Of course, the inclusion of two-loop diagrams might eventually change that. However, since they are neglected here, we take them into account indirectly by choosing appropriate parameters for the three-gluon vertex model. Thus we define what we call the optimized effective three-gluon vertex model as \eref{eq:3g-new} with the parameters chosen such that the propagators are reproduced as good in agreement with lattice data as possible. Note that other sources of deviations are neglected tensor structures of the three-gluon vertex or the truncation of the ghost-gluon vertex DSE. All these effects are included effectively in our choice of parameters, but without further studies we cannot say how much each one contributes.

\begin{figure}[tb]
 \begin{center}
  \includegraphics[width=0.49\textwidth]{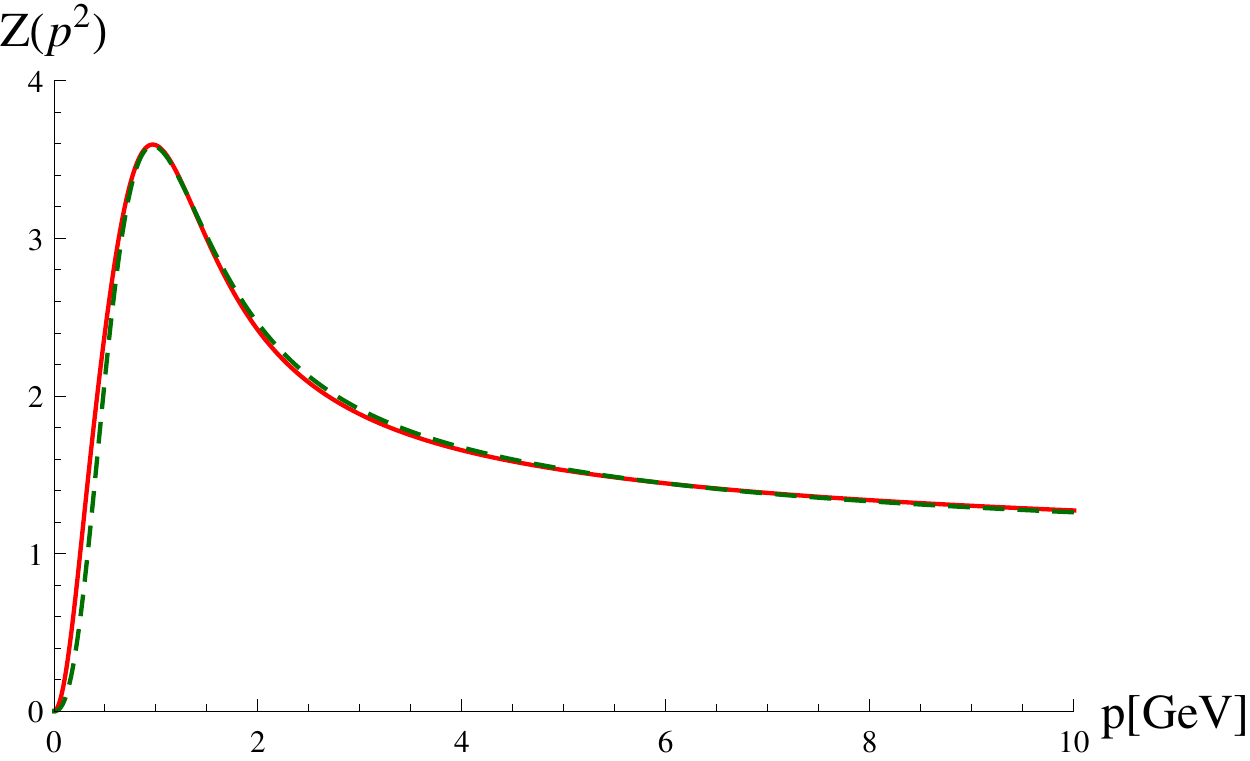}
  \includegraphics[width=0.49\textwidth]{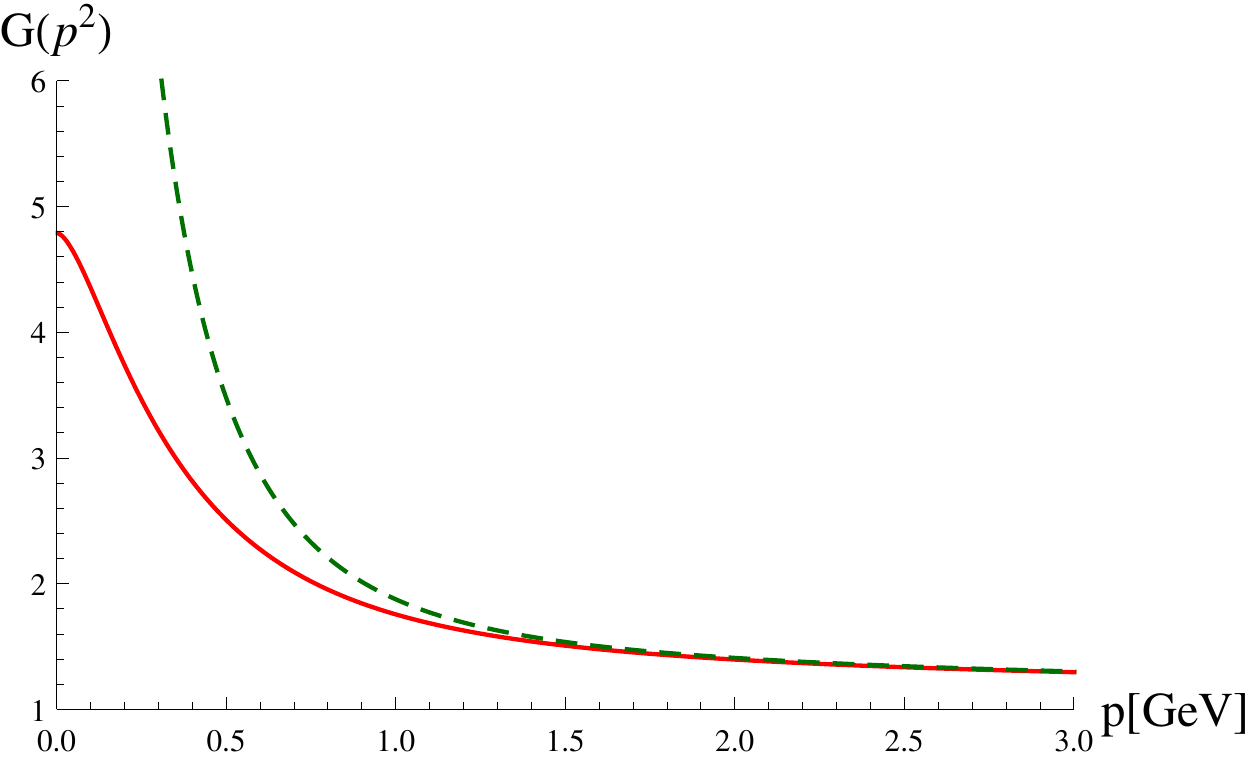}
  \caption{\label{fig:compDynGhgDecSca_props}The gluon propagator dressing function $Z(p^2)$ and the ghost dressing function $G(p^2)$ with a dynamic ghost-gluon vertex for decoupling (red/continuous line, same as in \fref{fig:compLDynGhg_props}) and scaling (green/dashed line). The parameter $\Lambda_{3g}$ for the latter is $2.05\,GeV$.}
 \end{center}
\end{figure}

We compare the resulting propagators to lattice results in \fref{fig:compLDynGhg_props}. For reference we also show the result from a propagator only calculation with the three-gluon vertex of \eref{eq:3g-Fischer}. Clearly, the gap between lattice and DSE results in the mid-momentum regime can be considerably diminished. The changes for the ghost dressing are small. For the gluon propagator the IR part looks overenhanced. However, this region is very sensitive to the scale setting, which was done here by placing the maximum of the gluon dressing at the same position as in the lattice data. This implies that we have fixed the scale to be the same as in the corresponding lattice simulations. There, the coupling defined by~\cite{vonSmekal:1997is,vonSmekal:1997vx}
\begin{align}\label{eq:coupling}
 \alpha(p^2)=\alpha(\mu^2)G(p^2)^2 Z(p^2),
\end{align}
where $\alpha(\mu^2)=g^2/4\pi$,
corresponds to $\Lambda^{N_f=0}_{\overline{MS}}$ of roughly $260\, MeV$ \cite{Sternbeck:2010xu,Sternbeck:2012qs}. The corresponding plot can be found in \fref{fig:compLDynGhg_props}. We obtain good agreement with the coupling from the lattice, for example we have $\alpha(178.5\,GeV^2)=0.143$ and $\alpha(1785\,GeV^2)=0.110$,\footnote{These momentum values correspond to $1000$ and $10000$ in units of the Sommer scale $r_0^2$.} which is in good agreement with the $N_f=0$ results of roughly $0.142(2)$ and $0.107(1)$, respectively \cite{Sternbeck:2012qs}. The Schwinger function of the gluon propagator, which shows clear signs of positivity violation, can be found in Ref. \cite{Huber:2013xb}. The corresponding scaling solution is shown in \fref{fig:compDynGhgDecSca_props}.

\begin{figure}[tb]
 \begin{center}
  \includegraphics[width=0.49\textwidth]{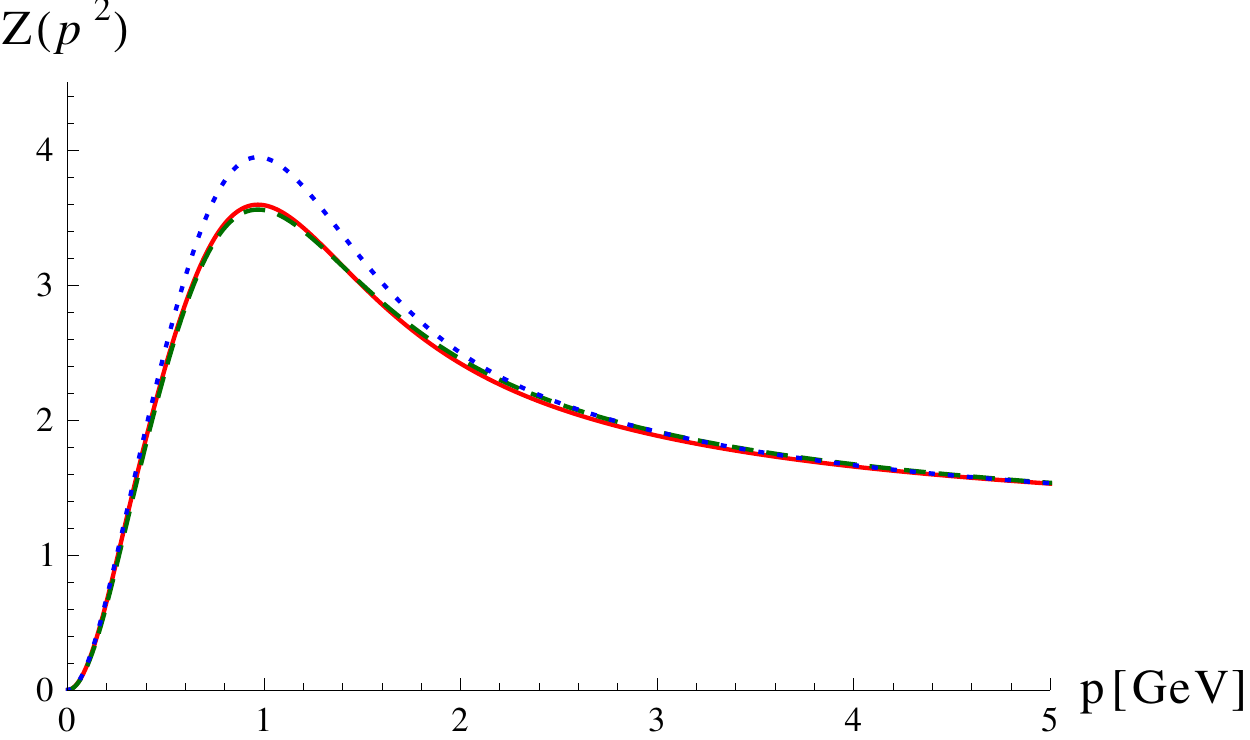}
  \includegraphics[width=0.49\textwidth]{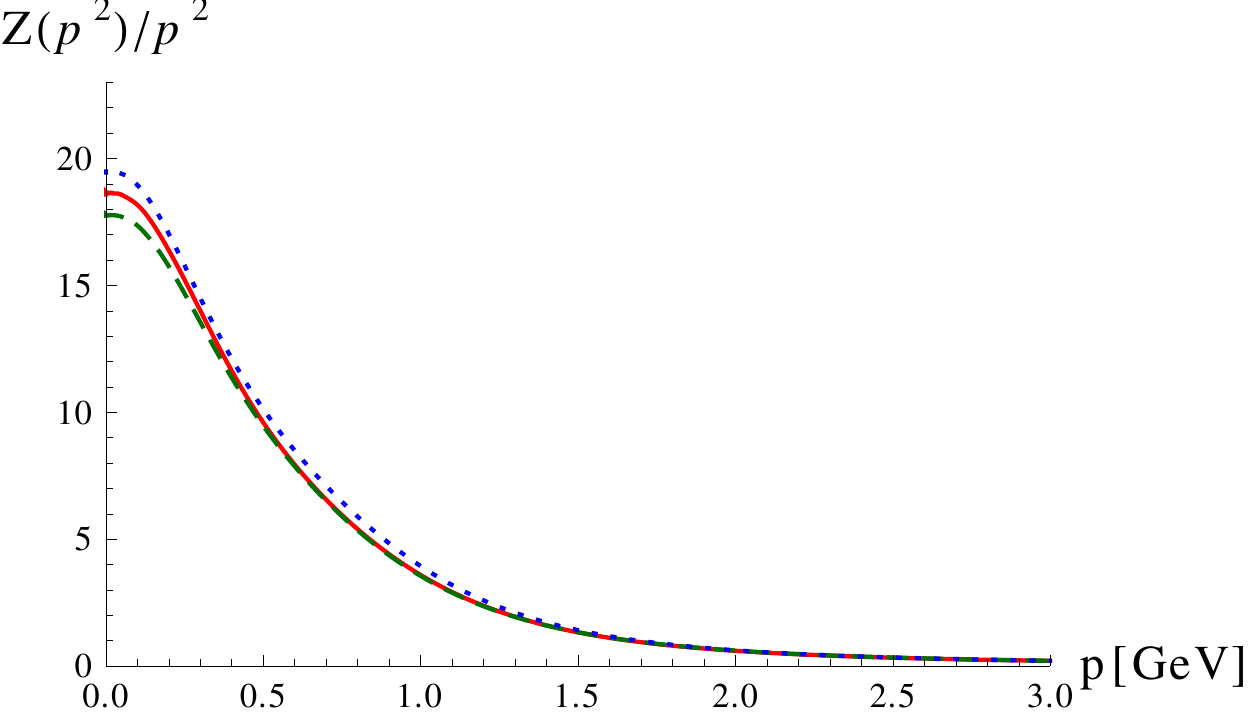}\\
  \includegraphics[width=0.49\textwidth]{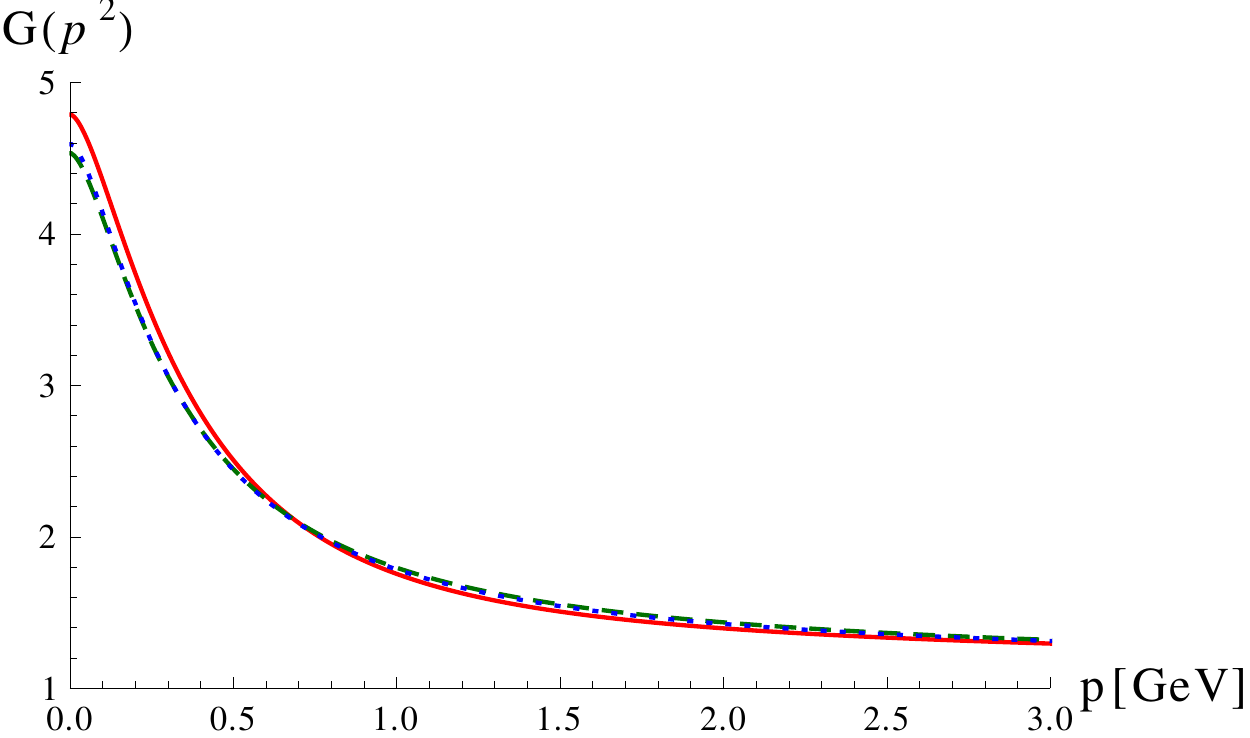}\hspace{0.5\textwidth}
  \caption{\label{fig:compOptEff3g_props}Comparison of the results for the gluon propagator dressing function $Z(p^2)$, the gluon propagator $Z(p^2)/p^2$ and the ghost dressing function $G(p^2)$. A dynamic (red/continuous line, parameter set 1) and a bare (green/dashed line, parameter set 2) ghost-gluon vertex are used together with the corresponding optimized effective three-gluon vertices.
The blue/dotted line (parameter set 3) stems from a calculation using the same parameters as in the dynamic calculation and a bare ghost-gluon vertex. See Tab.~\ref{tab:paras} for details on the parameter sets.}
 \end{center}
\end{figure}

Assessing the effect on the propagators of the ghost-gluon vertex alone is not straightforward, because the three-gluon vertex model depends on the propagators which are in turn changed by the ghost-gluon vertex. Thus even for the same parameters the model is not the same with or without a dynamic ghost-gluon vertex. The following comparison is therefore to be considered with a grain of salt. In \fref{fig:compOptEff3g_props} we show three solutions: The one with a dynamic ghost-gluon vertex and the optimized effective three-gluon vertex from above and two others with a bare ghost-gluon vertex. When using the same parameters with a bare ghost-gluon vertex, the mid-momentum enhancement of the gluon dressing function is too strong. Adjusting the parameters to the bare ghost-gluon vertex, however, shows that the three-gluon vertex model can incorporate even effects of the ghost-gluon vertex, as this solution then agrees very well with the one obtained from the dynamic ghost-gluon vertex. Furthermore one can also see that the ghost dressing function is almost unaffected by changing the three-gluon vertex parameters and that it is only slightly modified by the dynamic ghost-gluon vertex. The employed parameters for the three-gluon vertex model are summarized in Tab.~\ref{tab:paras}.

\begin{table}[tb]
\begin{center}
\begin{tabular}{c|c|c|c}
  parameter set & $h_{IR}$ & $\Lambda_{3g}$ & ghost-gluon vertex\\
  \hline\hline
  1 & $-1$ & $2.1\,GeV$ & dynamic\\
  \hline
  2 & $-1$ & $1.8\,GeV$ &  bare\\
  \hline
  3 & $-1$ & $2.1\,GeV$ & bare
\end{tabular}
\caption{\label{tab:paras}Used parameters for the three-gluon vertex model. Sets $1$ and $2$ correspond to the optimized effective three-gluon vertex model for the respective choice of the ghost-gluon vertex.}
\end{center}
\end{table}

\subsection{Ghost-gluon vertex}

\begin{figure}[tb]
 \begin{center}
 \begin{minipage}[t]{0.48\textwidth}
  \includegraphics[width=\textwidth]{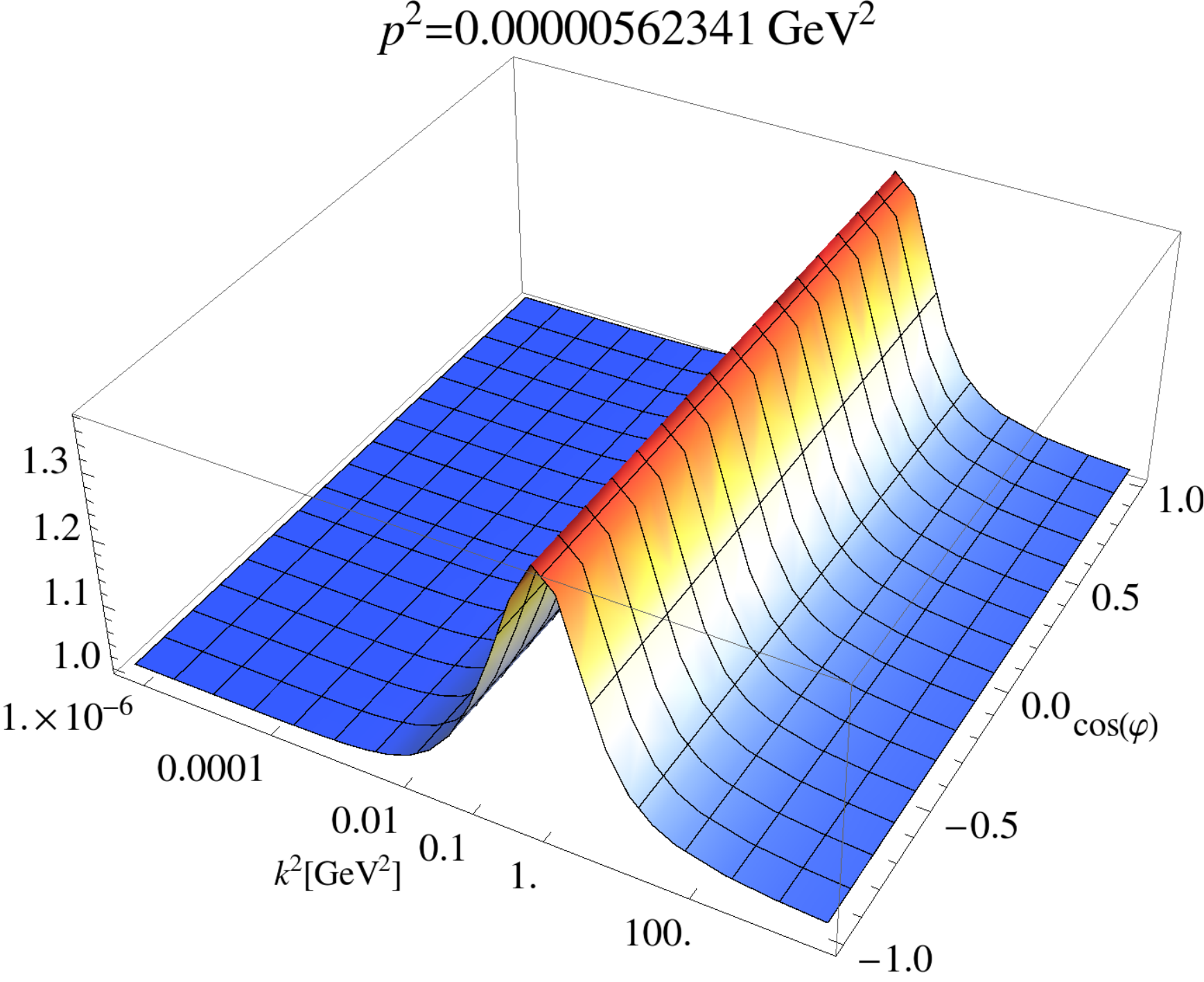}
 \end{minipage}
 \hfill
 \begin{minipage}[t]{0.48\textwidth}
  \includegraphics[width=\textwidth]{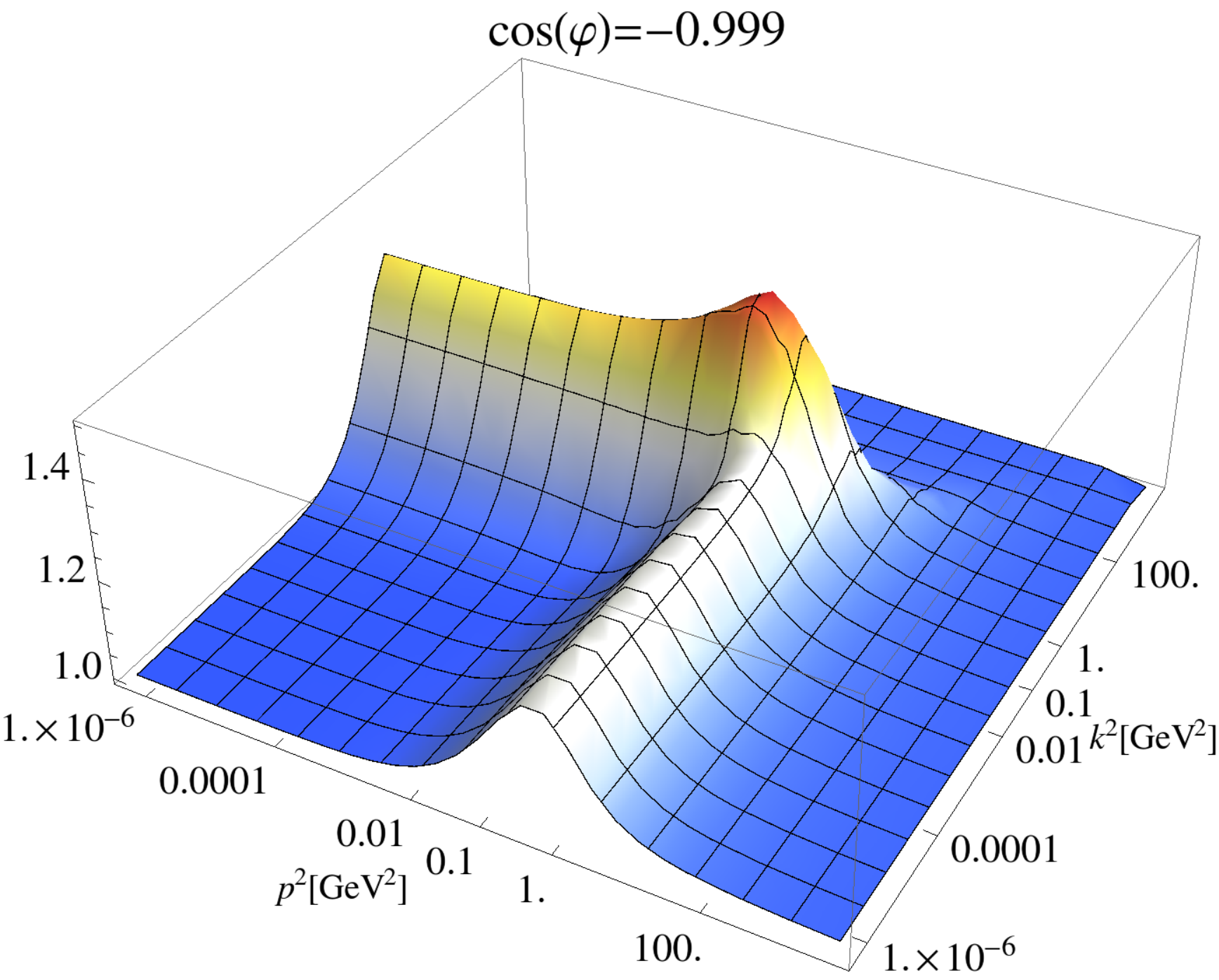} 
 \end{minipage}
  \begin{minipage}[t]{0.48\textwidth}
  \includegraphics[width=\textwidth]{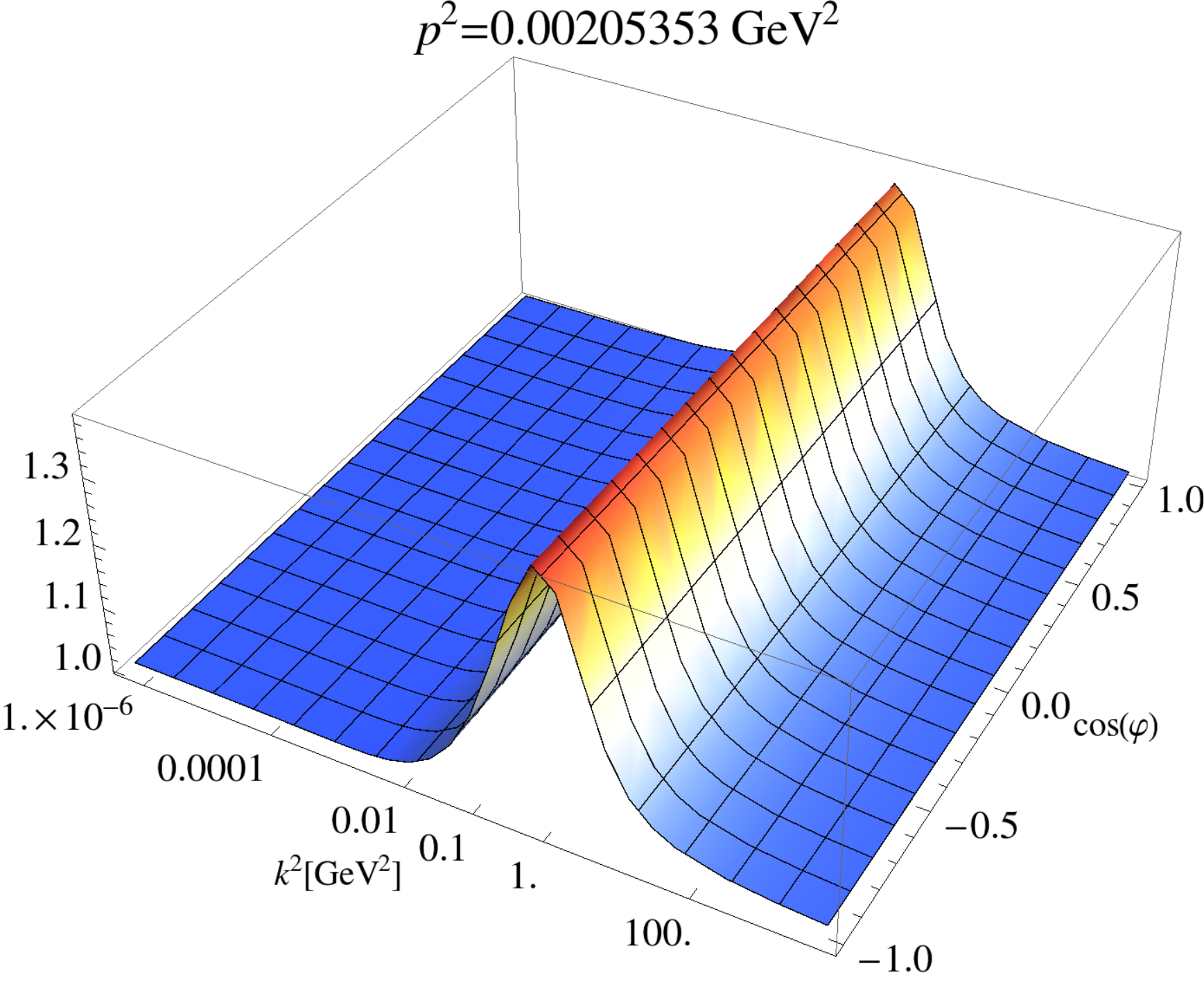}
 \end{minipage}
 \hfill
 \begin{minipage}[t]{0.48\textwidth}
  \includegraphics[width=\textwidth]{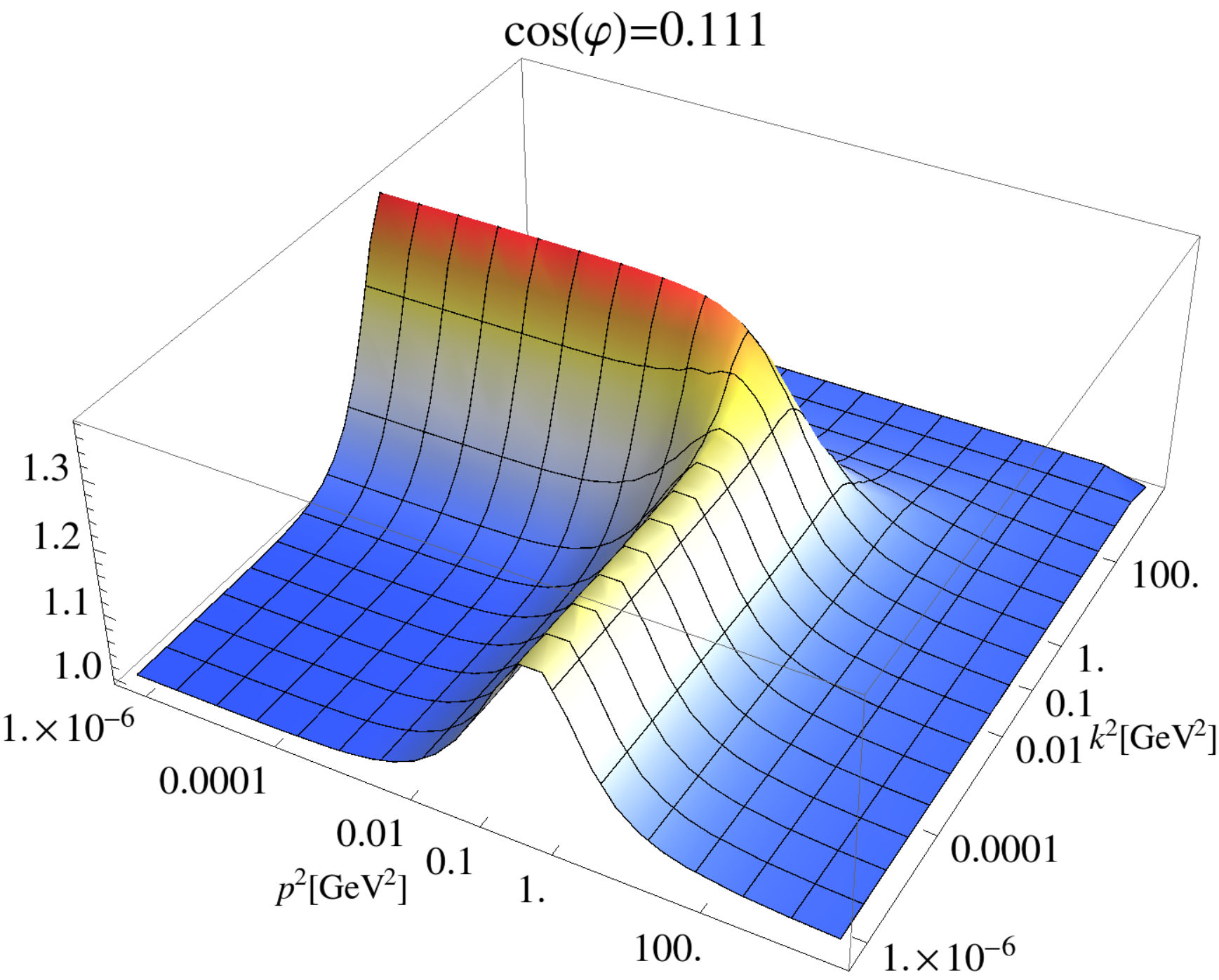} 
 \end{minipage}
 \caption{\label{fig:dynGhgDec_ghg}Selected momentum configurations of the ghost-gluon vertex for a decoupling solution. Fixed momentum (\textit{left}) or angle (\textit{right}) as indicated at the top of the plots.}
 \end{center}
\end{figure}

\begin{figure}[tb]
 \begin{center}
  \includegraphics[width=0.49\textwidth]{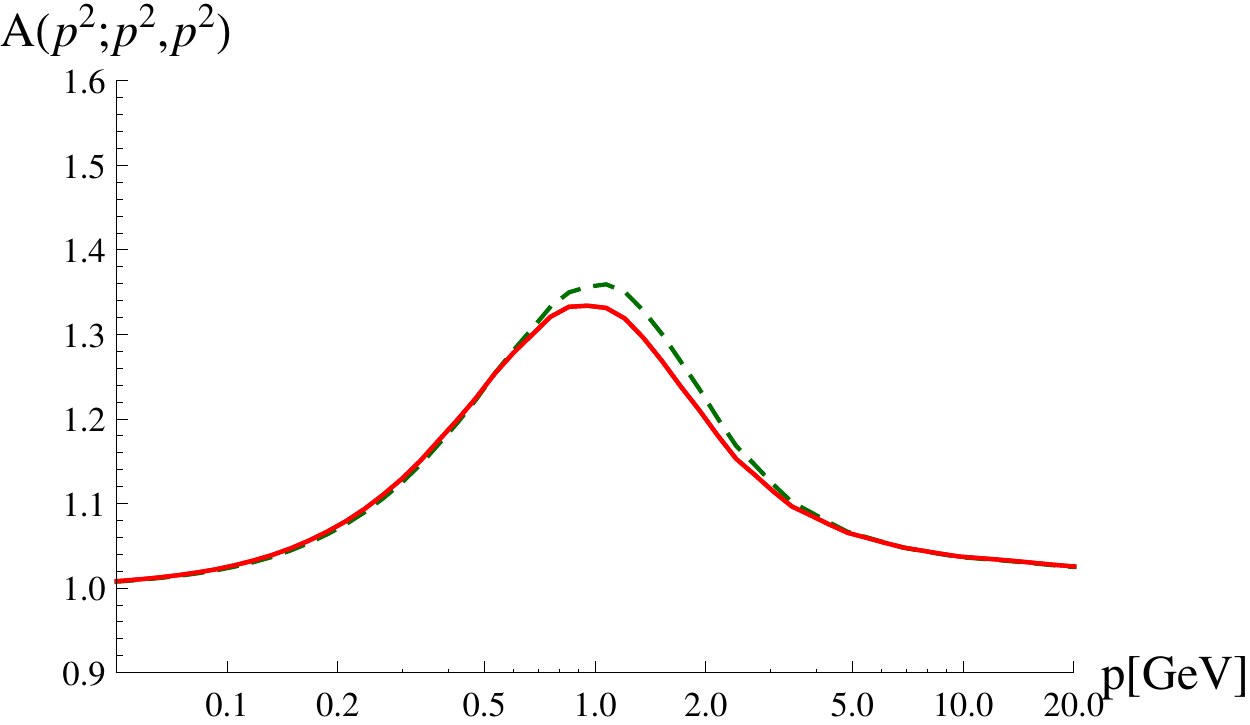}
  \includegraphics[width=0.49\textwidth]{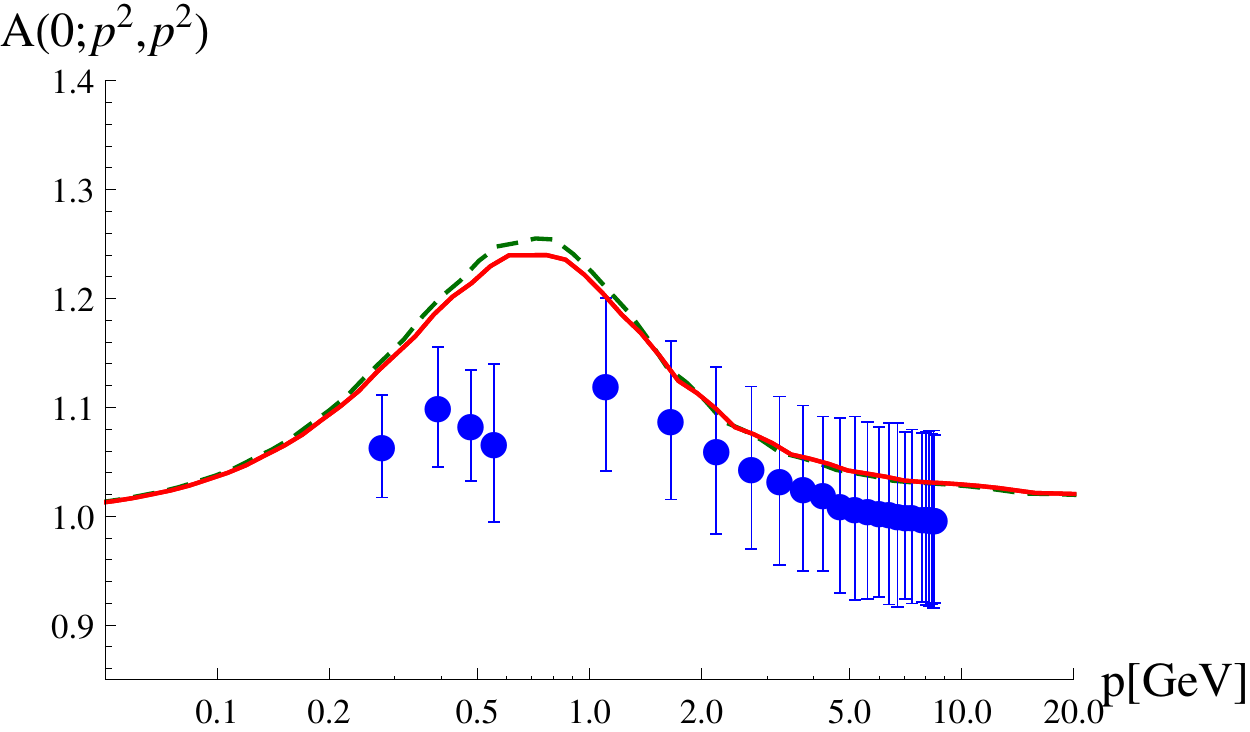}
  \caption{\label{fig:dynGhgDec_ghg_1scale}The ghost-gluon vertex for one-scale momentum configurations. The red continuous line is the solution with the optimized effective three-gluon vertex ($h_{IR}=-1$ and $\Lambda_{3g}=2.1\,GeV$), the green dashed one was obtained with the same $h_{IR}$ but for $\Lambda_{3g}=2.6\,GeV$. \textit{Left:} Symmetric momentum configuration $p^2=q^2=k^2$. \textit{Right:} The configuration with vanishing gluon momentum in comparison with lattice results 
from Ref.~\cite{Sternbeck:2006rd} (blue circles: $N=48$ at $\beta=6$).}
 \end{center}
\end{figure}

Next we will present results for the ghost-gluon vertex.
Up to now numerical non-perturbative investigations of the ghost-gluon vertex were only done for specific momentum configurations \cite{Schleifenbaum:2004id,Cucchieri:2008qm} or were restricted to the IR part \cite{Alkofer:2008dt}. In Ref.~\cite{Dudal:2012zx} a ghost-gluon vertex model based on an operator product expansion \cite{Boucaud:2011eh} was adapted such as to properly describe the ghost dressing via its DSE.
Qualitatively the vertex model with the determined parameters resembles lattice results at least for the momentum configuration of vanishing ghost momentum. It also contains a peak at intermediate momenta which is, however, somewhat higher than in Monte Carlo simulations and here. Moreover, it should be noted that the choice of parameters is not necessarily unique and while it leads to good results for the ghost propagator, it might not be an adequate description, for instance, in the gluon DSE. In addition, the model of Ref.~\cite{Dudal:2012zx} simplifies the original proposal in Ref.~\cite{Boucaud:2011eh} by setting the ghost momentum to zero what should affect the parameters. Here we follow another approach and obtain the complete momentum dependence from a self-consistent calculation.
Thereby it turns out that the structure is in general as simple as for special momentum configurations: The vertex is one for large momenta, develops a small peak in the intermediate momentum regime and then settles for a constant value in the IR. To illustrate this we show a few exemplary momentum configurations in \fref{fig:dynGhgDec_ghg}. Additionally typical one-scale configurations are shown in \fref{fig:dynGhgDec_ghg_1scale}. For the so-called asymmetric momentum configuration, where the gluon momentum vanishes, we compare our result with lattice data. More lattice results with better statistics 
exist for $SU(2)$ \cite{Cucchieri:2008qm}. However, in this $SU(2)$ data the peak of the vertex dressing occurs at somewhat larger momenta than in our calculations. Whether this is a genuine difference between $SU(2)$ and $SU(3)$, whether it indicates an inconsistency in the scale setting or whether the effect might be due to our truncation employed for the vertex DSE remains to be further investigated.

In addition to the ghost-gluon vertex obtained with the optimized effective three-gluon vertex model, we show in \fref{fig:dynGhgDec_ghg_1scale} for comparison also a result obtained with the same $h_{IR}$ but a considerably larger $\Lambda_{3g}$. Both results yield a very similar ghost-gluon vertex. Therefore we conclude that the vertex is rather insensitive to the three-gluon vertex model in our truncation, although this could be expected as an indirect effect via the gluon propagator. If we proceeded one step further in the truncation of the ghost-gluon vertex DSE, by including the dressed three-gluon vertex, this will most likely change. If in a next step the three-gluon vertex is calculated dynamically this should yield a viable improvement of the ghost-gluon vertex truncation in the future.

\begin{figure}[tb]
 \begin{center}
 \begin{minipage}[t]{0.48\textwidth}
  \includegraphics[width=\textwidth]{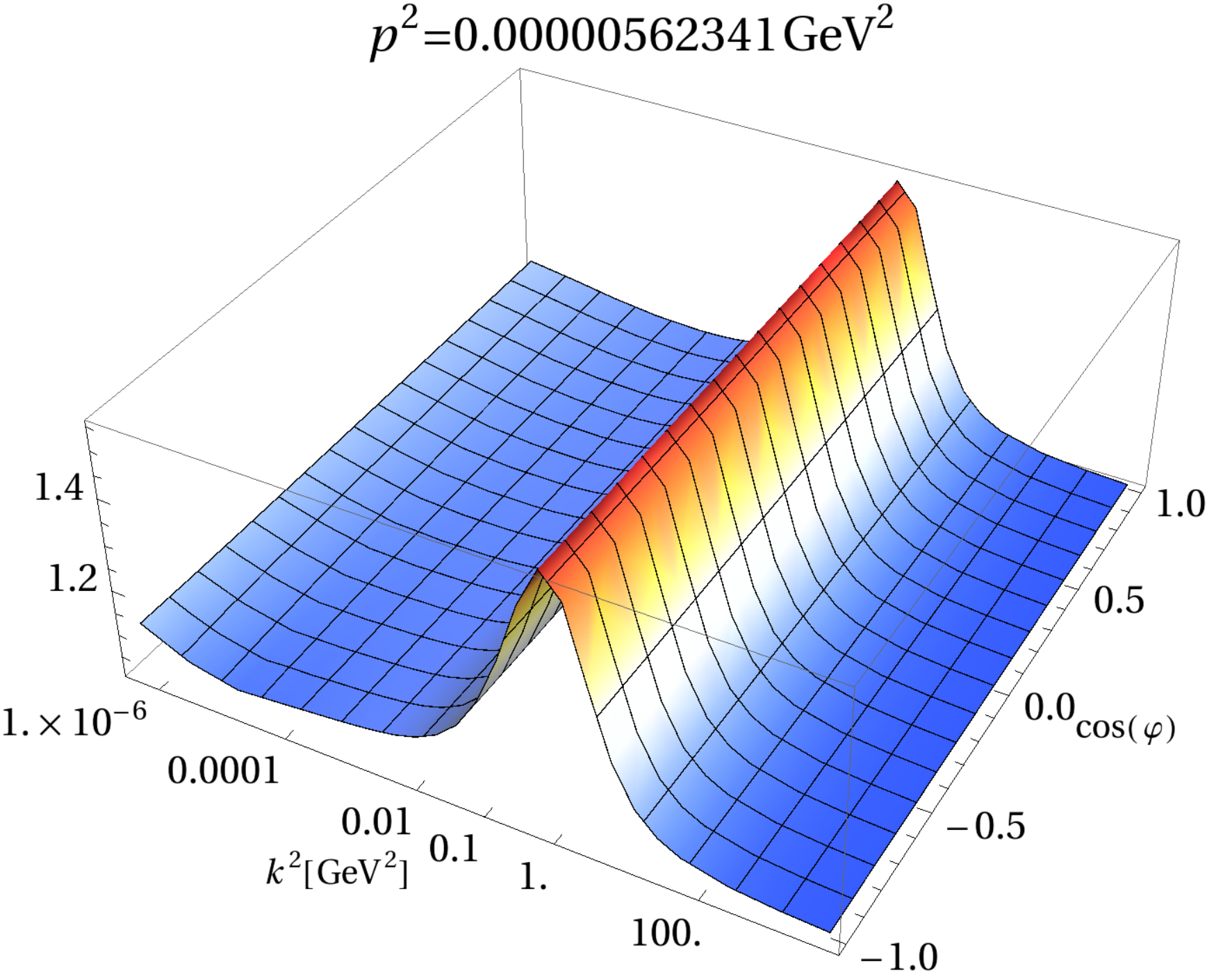}
 \end{minipage}
 \hfill
 \begin{minipage}[t]{0.48\textwidth}
  \includegraphics[width=\textwidth]{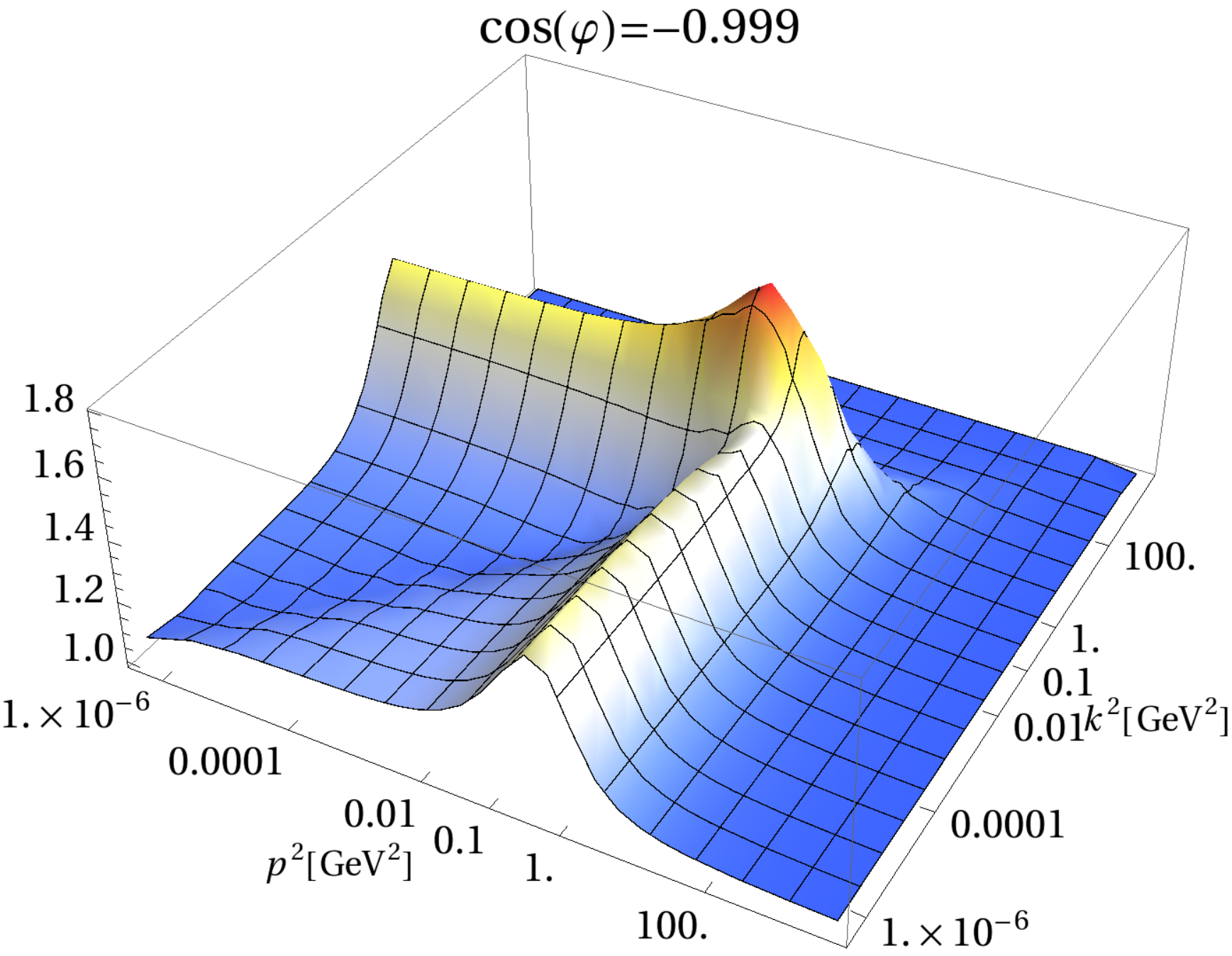} 
 \end{minipage}
  \begin{minipage}[t]{0.48\textwidth}
  \includegraphics[width=\textwidth]{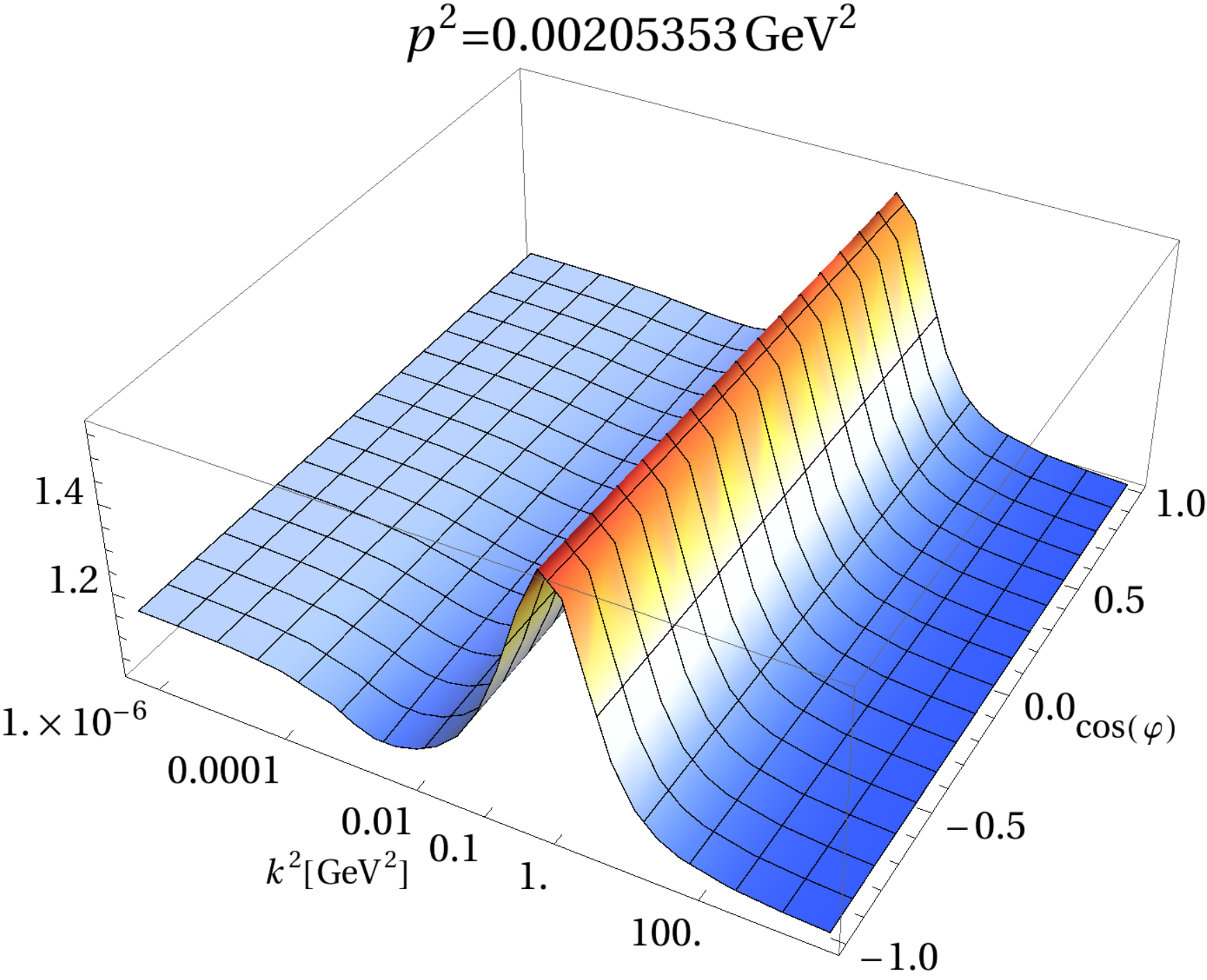}
 \end{minipage}
 \hfill
 \begin{minipage}[t]{0.48\textwidth}
  \includegraphics[width=\textwidth]{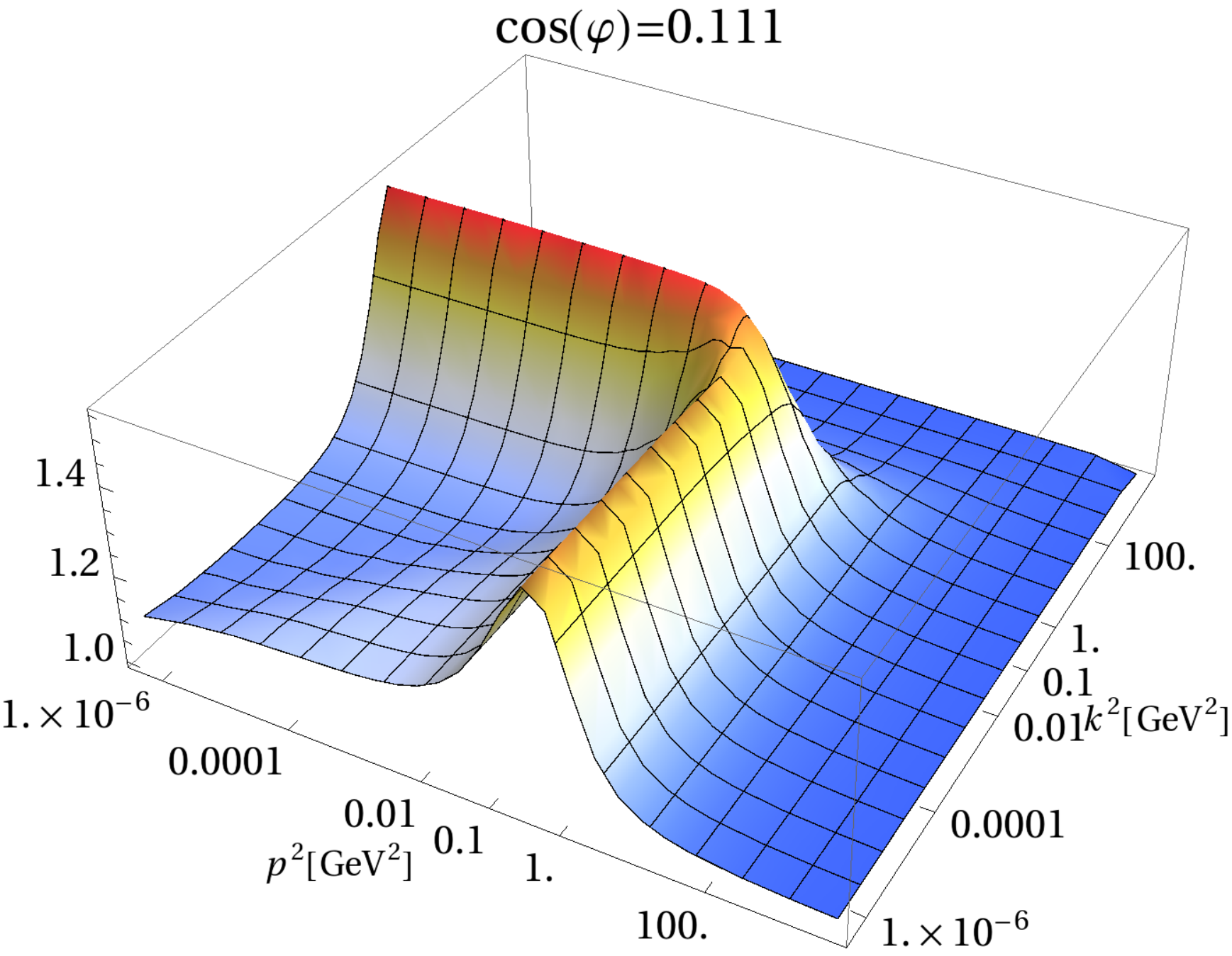} 
 \end{minipage}
 \caption{\label{fig:dynGhgSca_ghg}Selected momentum configurations of the ghost-gluon vertex for the scaling solution. Fixed momentum (\textit{left}) or angle (\textit{right}) as indicated at the top of the plots.}
 \end{center}
\end{figure}

The system of equations can also be solved for scaling boundary conditions, viz. $1/G(0)=0$ \cite{Lerche:2002ep}. By varying the boundary condition to a finite value a family of decoupling solutions can be obtained \cite{Boucaud:2008ji,Fischer:2008uz,Alkofer:2008jy}. Above this value was chosen such to reproduce the lattice results for the ghost dressing function. Up to now it has not been investigated numerically if this ambiguity in the solutions also persists for the vertices. Qualitatively no difference is expected for the ghost-gluon vertex \cite{Alkofer:2008jy}: It should become IR finite for both types of solutions. Any difference must therefore be of quantitative nature. Indeed we find such a difference as illustrated in Figs.~\ref{fig:dynGhgDec_ghg} and \ref{fig:dynGhgSca_ghg} for selected momentum configurations: For scaling the IR value of the ghost-gluon vertex is larger than one, but for decoupling it is exactly one, i.e., all diagrams vanish in this limit. This can be understood from IR power counting of the dressing functions: In the Abelian diagram we have two ghost propagators and one gluon propagator. We get in the limit that all external momenta approach zero with one scale that the diagram scales as $p^2$ for decoupling, i.e., it vanishes, while for scaling the IR exponent is $-2\ka+2\ka=0$. Thus the diagram is IR finite. The non-Abelian diagram with a bare three-gluon vertex is IR suppressed for both types of solutions. It is important to note that our result of a ghost-gluon vertex not approaching one in the IR is not in conflict with Taylor's theorem, since it only concerns the longitudinal dressing function as discussed in Section~\ref{sec:ghg}.

\section{Summary and outlook}
\label{sec:summary}

In the present work we have extended typical truncation schemes of DSEs by including the ghost-gluon vertex dynamically together with the propagators in the calculations. The employed DSE truncations contain the IR and UV leading diagrams. The results for the vertex are in good agreement with available lattice data: Basically it stays flat with a small bump around $1\,GeV$. The influence on the propagators is most visible in the mid-momentum regime. As far as the different types of solutions for the propagators are concerned, this ambiguity is also present  at the level of the vertex: For decoupling the vertex approaches one in the IR, while a scaling vertex receives IR contributions from the IR leading diagram thus settling at a value larger than one for asymptotically small momenta.

The second improvement concerns the three-gluon vertex: While in many truncation schemes it is adjusted so as to obtain the correct UV anomalous dimension for the gluon propagator, we have devised a model that respects Bose symmetry and qualitatively captures the behavior seen in lattice calculations, especially the fact that its dressing function turns negative in the IR. Since only two- and three-dimensional lattice data is unequivocal in this respect, we have calculated the IR leading contribution of the three-gluon vertex at the symmetric point. We did find a zero crossing, but at momenta lower than the lowest available momentum in lattice simulations. Thus the zero crossing itself is not relevant for the gluon DSE as all contributions in this regime get suppressed by the two attached gluon propagators. This is also corroborated by an analysis of the influence of the IR part in our model. However, the Bose symmetrization of the vertex leads to an increase of the mid-momentum strength of the gluon propagator. By choosing the parameters of the model appropriately, we could effectively incorporate the effects of two-loop diagrams in the mid-momentum regime. With this optimized effective three-gluon vertex we were able to close the gap to lattice data.

An important next step in extending the truncation scheme is certainly the three-gluon vertex. If included self-consistently one can then disentangle contributions due to two-loop diagrams from the gluon loop. For the former the three-gluon vertex will also be required as it is contained in the so-called squint diagram. Furthermore, an extension of the truncation of the ghost-gluon vertex DSE could be to use the DSE with the bare vertex attached to the ghost leg. This then includes the three-gluon vertex as well. As an alternative to DSEs it might be interesting to look also at the ghost-gluon vertex equation of motion from an nPI effective action, where the bare vertices get dressed. This might compensate for the neglected diagrams in the present truncation.

In summary, the presented extension of a standard truncation scheme confirms the qualitative reliability of previous schemes. Furthermore, it shows that by including higher vertex functions DSEs can successfully be used to obtain quantitative results, although the underlying system of equations is infinitely large. Thus DSEs provide a rather satisfying description of the non-perturbative regime which can be systematically improved.

\section*{Acknowledgments}
We thank Reinhard Alkofer, Christian~S. Fischer, Leonard Fister, Valentin Mader, Mario Mitter, Jan M.~Pawlowski and Stefan Strauss for useful discussions. We are also grateful to  Axel Maas and Andr\'e Sternbeck for providing lattice results and valuable discussions.
M.Q.H.\ was supported by the Alexander von Humboldt foundation, L.v.S.\ by the Helmholtz International Center for FAIR within the LOEWE program of the State of Hesse, the Helmholtz Association Grant VH-NG-332, and the European Commission, FP7-PEOPLE-2009-RG No.~249203. Plots of DSEs were created with \textit{FeynDiagram} and \textit{JaxoDraw} \cite{Binosi:2003yf} and other images with \textit{Mathematica} \cite{Wolfram:2004}.

\appendix

\section{Kernels}
\label{sec:app_kernels}

The kernel $K_{G}$ of the ghost DSE given in \eref{eq:gh-DSE} is
\begin{align}
 K_{G}(p,q)=\frac{ x^2+(y-z)^2-2 x (y+z)}{4 x y^2 z}
\end{align}
with $x=p^2$, $y=q^2$ and $z=(p+q)^2$.
The two kernels $K_{Z}^{gh}$ and $K_{Z}^{gl}$ of the gluon DSE \eref{eq:gl-DSE} read
\begin{align}
 K_{Z}^{gh}(p,q)&=-\frac{ x^2+(y-z)^2-2 x (y+z)}{12 x^2 y z},\\
 K_{Z}^{gl}(p,q)&=\frac{1}{24 x^2 y^2 z^2}\Big(x^4+8 x^3 (y+z)+x^2 \left(-18 y^2-32 y z-18 z^2\right)+\nnnl
  &(y-z)^2 \left(y^2+10 y z+z^2\right)+2x(y+z)\left(4y^2-20y z+4z^2\right)\Big).
\end{align}
The appearance of spurious quadratic divergences is discussed in Appendix~\ref{sec:props_UV}. The gist is that they are subtracted via additional terms that contain a damping function such the IR behavior is not influenced. For the respective damping scales in the ghost and gluon loops we used $L^2_{UV}=10$ and $2$ in internal units for the ghost and gluon loops. The physical values can only be determined at the end of the calculation when the scale is set.

\begin{figure}
 \begin{center}
  \includegraphics[width=0.49\textwidth]{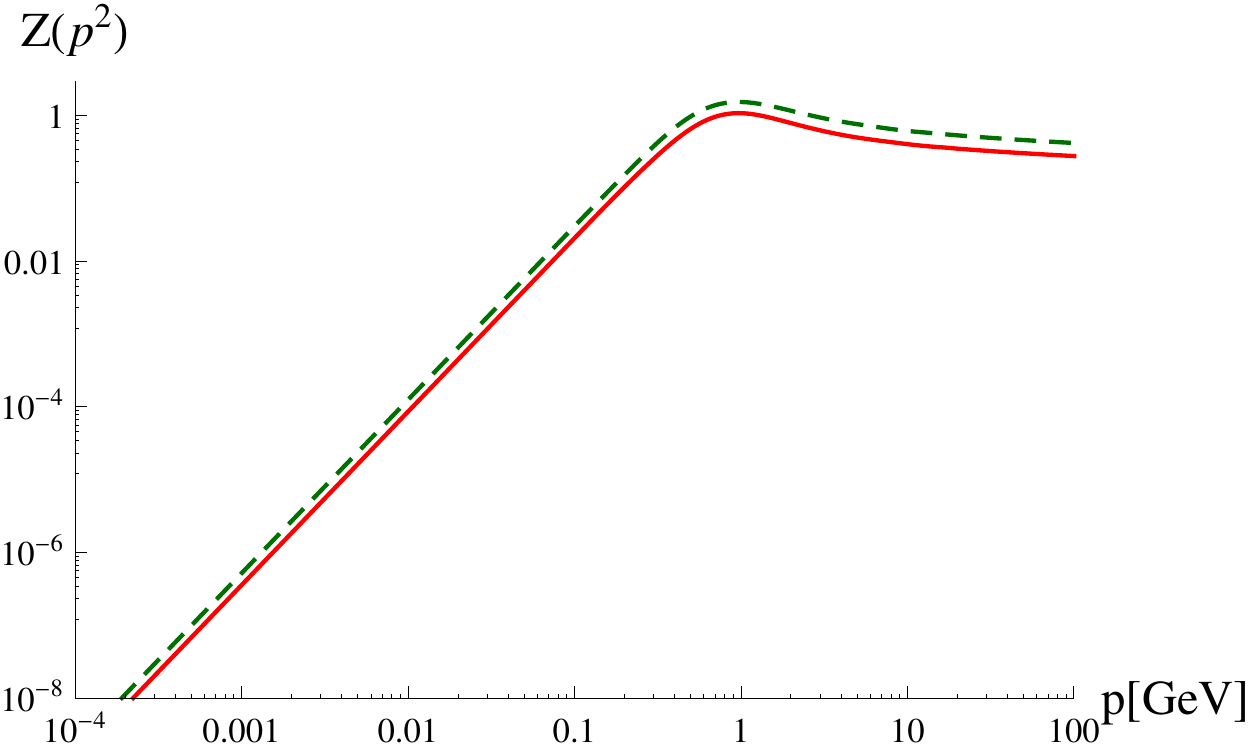}
  \includegraphics[width=0.49\textwidth]{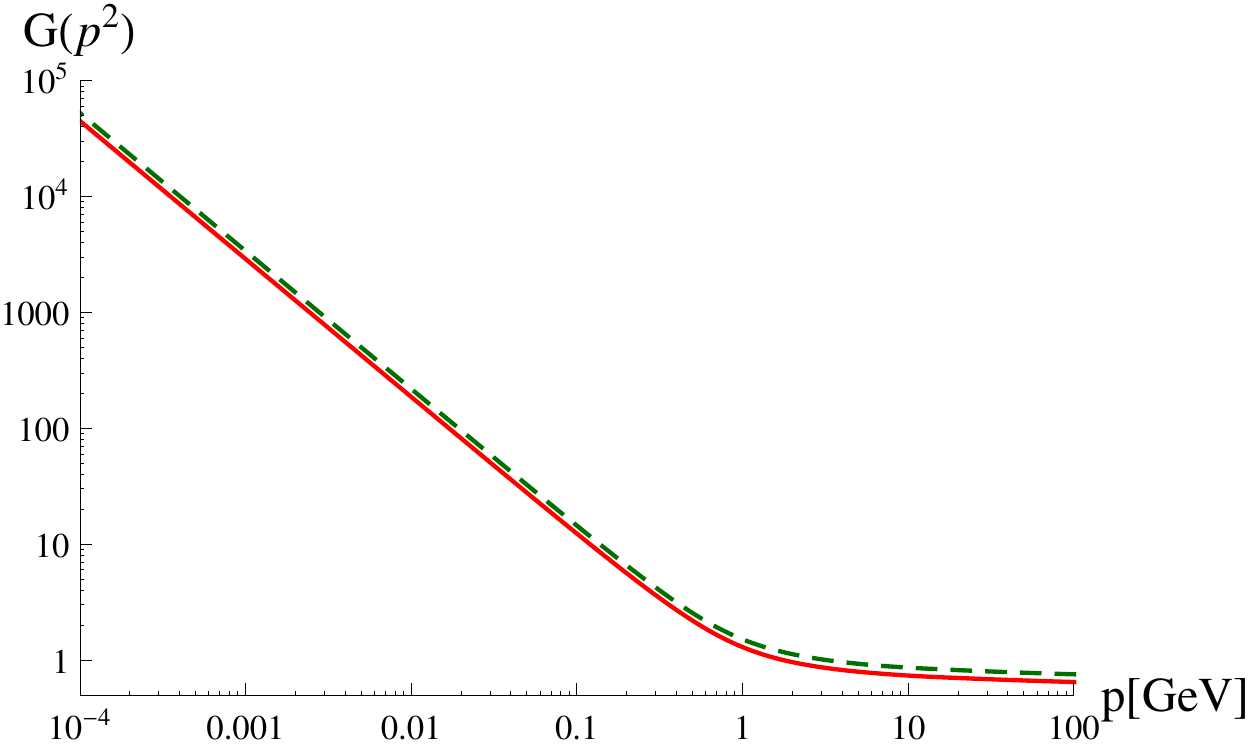}\\
  \includegraphics[width=0.49\textwidth]{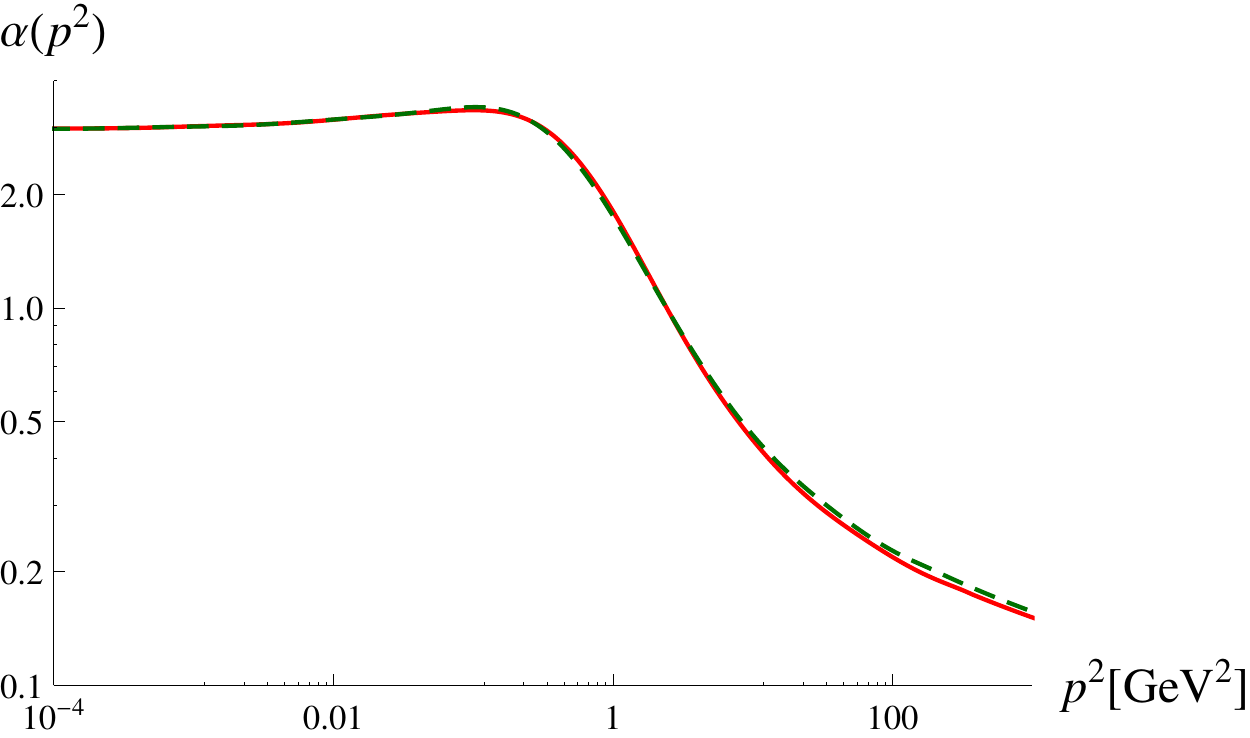}\hspace{0.5\textwidth}
  \caption{\label{fig:compMultRenorm}Test of multiplicative renormalizability, here for the scaling solution: The gluon and ghost dressing functions (\textit{top left} and \textit{right}, respectively) depend on the renormalization point, which is indirectly fixed by the choice of $\alpha(\mu^2)$. The red/continuous curve is for $\alpha(\mu^2)=1$, the green/dashed one for $\alpha(\mu'^2)=0.5$. The running coupling (\textit{bottom}) is renormalization group invariant and thus does not change. The small deviation in the UV is a numeric artifact: The physical values for $L_{UV}$ do not match exactly, because the scale is set at the end and we can only get approximately equal values.}
 \end{center}
\end{figure}

Untruncated DSEs respect multiplicative renormalizability. The employed truncation and the method for subtracting the quadratic divergences do not change this as illustrated in \fref{fig:compMultRenorm}. There we show the results for the dressings and the running coupling if different values for $\alpha(\mu^2)$ are chosen. This indirectly fixes the renormalization point $\mu^2$. For the comparison the values of the damping scale parameters $L_{UV}$ must have the same physical values, which can only be determined a posteriori.

\begin{figure}
 \begin{center}
  \includegraphics[width=0.49\textwidth]{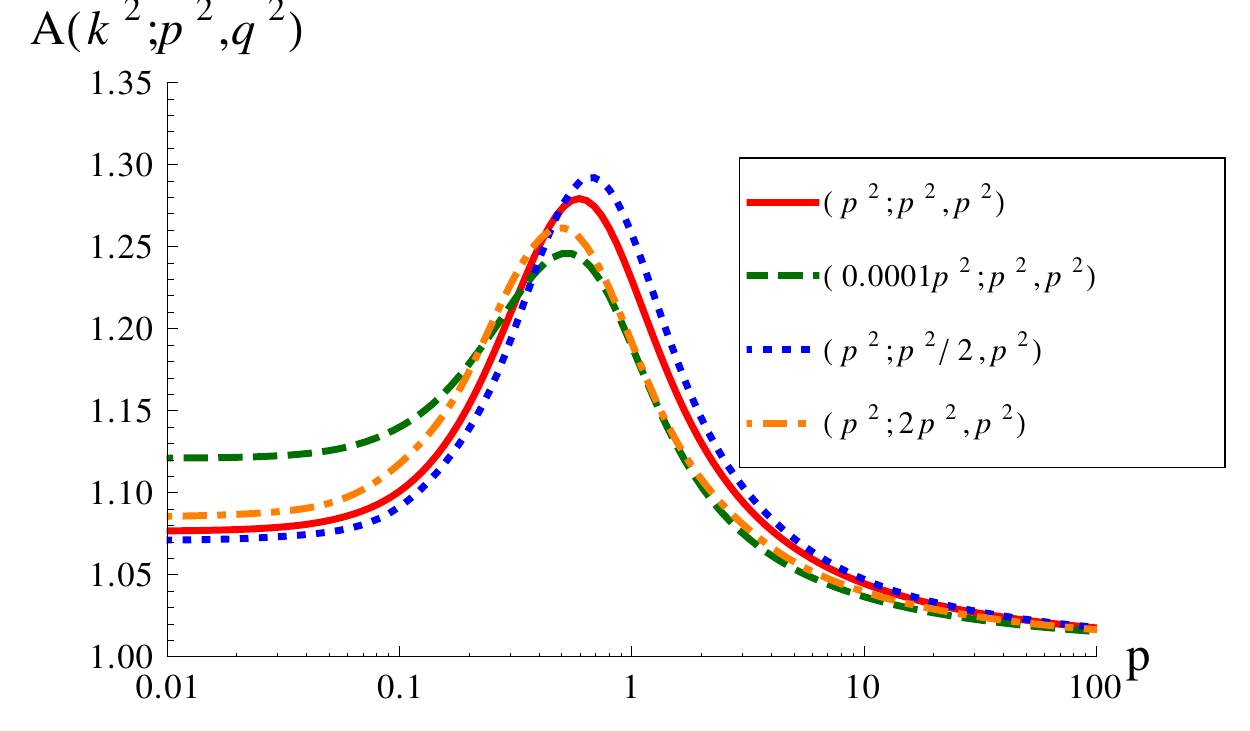}
  \includegraphics[width=0.49\textwidth]{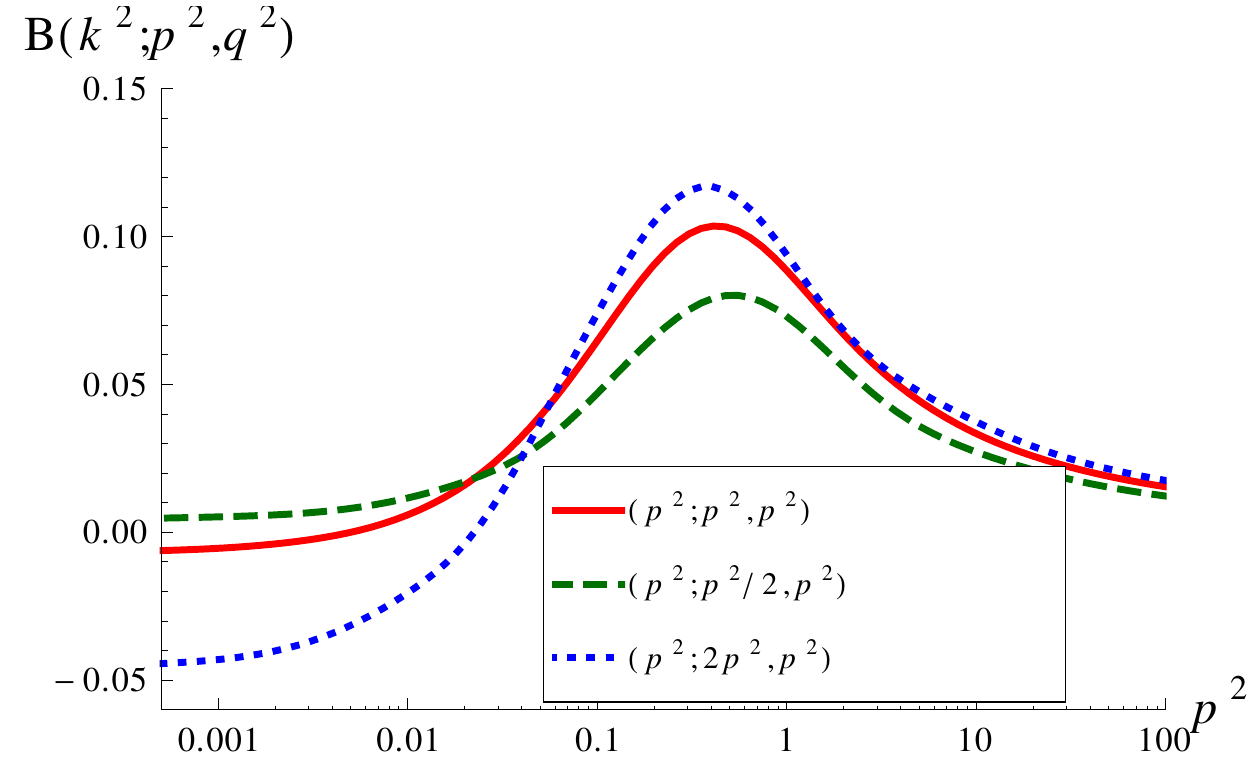}
  \caption{\label{fig:ghg_Schleifenbaum} Dressing functions $A(k;p,q)$ and $B(k;p,q)$ after one iteration for the same momentum configurations and the same analytic expressions for the propagators as in Refs.~\cite{Schleifenbaum:2004id} from our kernels, i.e., with the signs of the kernels of the Abelian diagram changed. The main differences arise in the IR, where the Abelian diagram is dominant.}
 \end{center}
\end{figure}

The kernels of the ghost-gluon vertex DSE \eref{eq:ghg-DSE} read:
\begin{align}
 K^{A\bar{c}c}_1&=\frac{(k\cdot r \, p \cdot k - k^2 p \cdot r) ((p \cdot r)^2 + k \cdot r (-p^2 + p \cdot r) - p^2 r^2 + 
   p \cdot k (p \cdot r - r^2))}{2 (p - r)^4 r^2 (k + r)^2},\\
 K^{A\bar{c}c}_2&=\frac{1}{(p - r)^2 r^4 (k + r)^4}\Big((k \cdot r\, p \cdot k - k^2\, p \cdot r) ((k \cdot r)^2 p \cdot r + 
     r^2 (p \cdot k\, p \cdot r + (p \cdot r)^2 - p^2 r^2 - p \cdot k\, r^2) \nnnl 
  &+     k^2 (k \cdot r\, p \cdot r - (p^2 + p \cdot k - p \cdot r) r^2) + 
     k \cdot r (p \cdot k\, p \cdot r + (p \cdot r)^2 - 2 p^2 r^2 - 2 p \cdot k\, r^2 + 
        p \cdot r\, r^2)) \nnnl
 &+ (k^2 p^2 - 
     p \cdot k^2) (k \cdot r (2 k \cdot r\, p \cdot r - k \cdot r\, r^2 - p \cdot k\, r^2) + 
     k^2 (k \cdot r\, p \cdot r + r^2 (-p \cdot k - p \cdot r + r^2)))\Big).
\end{align}
They were generated automatically using the program \textit{DoFun} \cite{Alkofer:2008nt,Huber:2011qr} and implemented in \textit{C++} with {\it CrasyDSE} \cite{Huber:2011qr}. The kernels given here and those for $B(k;p,q)$ were compared with those given in Ref.~\cite{Schleifenbaum:2004dt}. Except for a minus sign in front of the Abelian diagram the expressions agree. We repeated the calculation of Ref.~\cite{Schleifenbaum:2004dt} with our kernels. The results are shown in \fref{fig:ghg_Schleifenbaum}. Note that the different sign leads to an IR value which also differs from that in Ref.~\cite{Schleifenbaum:2004dt}.

\section{UV behavior of the propagator Dyson-Schwinger equations}
\label{sec:props_UV}

In the asymptotic UV regime the propagator dressing functions have a logarithmic behavior:
\begin{align}
 G(x)&=G(s)\left(\omega \log\left(\frac{x}{s}\right)+1 \right)^\delta,\label{eq:G-UV}\\
 Z(x)&=Z(s)\left(\omega \log\left(\frac{x}{s}\right)+1 \right)^\gamma,\label{eq:Z-UV}
\end{align}
where $\delta$ and $\gamma$ are the anomalous dimensions and $s$ is some high UV momentum. If both one loop diagrams are taken into account, $\omega$ is $11N_c \alpha(s)/12\pi$, $\delta=-9/44$ and $\gamma=-13/22$ \cite{Fischer:2002eq,Fischer:2003zc}. We will show how these values emerge from the propagator DSEs with the model for the three-gluon vertex.

We follow the calculation outlined in Ref. \cite{Atkinson:1997tu}, where the anomalous dimensions for the ghost-loop only truncation were calculated using the Brown-Pennington projector \cite{Brown:1988bn} for the gluon equation. For illustration purposes we use here as in Refs.~\cite{Fischer:2002eq,Fischer:2003zc} a generalized projector:
\begin{align}
 P^{\zeta}_{\mu\nu}(p)=g_{\mu\nu}-\zeta\frac{p_\mu p_\nu}{p^2}.
\end{align}
At the end we set $\zeta=1$ to recover the transverse projector.
Acting with this projector onto the gluon DSE yields:
\begin{align}
 \frac{1}{Z(p^2)}&=\tilde{Z}_3+N_c\,g^2\,\int_q G(q^2)G((p+q)^2) K_{Z}^{gh,\zeta}(p,q)\nnnl
 &+N_c\,g^2\,\int_q Z(q^2)Z((p+q)^2) K_{Z}^{gl,\zeta}(p,q)\Gamma^{A^3}(p,q,-p-q)
\end{align}
with the new kernels
\begin{align}
 K_{Z}^{gh,\zeta}(p,q)&=\frac{ 2 x (y+z)+x^2 (-2+\zeta )-(y-z)^2 \zeta }{12 x^2 y z},\\
 K_{Z}^{gl,\zeta}(p,q)&=\frac{z^2 \zeta }{24 x^2 y^2}+\frac{z (5 x-x \zeta +4 y \zeta )}{12 x^2 y^2}+\frac{x^2 (-19+\zeta )+2 x y (-17+\zeta )-18 y^2 \zeta }{24 x^2 y^2}\nnnl
 &+\frac{(x-y)^2 \left(x^2+10 x y+y^2 \zeta \right)}{24 x^2 y^2 z^2}+\frac{4 x^3+x y^2 (-17+\zeta )+4 y^3 \zeta -x^2 y (15+\zeta )}{12 x^2 y^2 z}
\end{align}
with $x=p^2$, $y=q^2$ and $z=(p+q^2)$.
As $P^\zeta$ also projects onto the longitudinal part of the ghost-gluon vertex, we already replaced it at this point by the bare expression.

Here we are interested only in high external momenta $p^2$, for which the integral is dominated by the region $x<y<L^2$, where $L$ is a $UV$ cutoff. Since the dressing functions $G(p^2)$ and $Z(p^2)$ vary only slowly for high momenta, we replace $(p+q)^2$ by $q^2$.
This allows to calculate the angle integrals analytically:
\begin{align}
 \frac{1}{G(p^2)}&\rightarrow Z_3+\frac{N_c\,g^2}{64\pi^2}\,\int_x^{L^2} dy \frac{x-3y}{ y^2} Z(y)G(y) \\
 \frac{1}{Z(p^2)}&\rightarrow \tilde{Z}_3+\frac{N_c\,g^2}{192\pi^2}\,\int_x^{L^2} dy \frac{-y(\zeta-4)+x(\zeta-2)}{ x\,y} G(y)G(y)\nnnl
& +\frac{N_c g^2}{384 \pi^2}\int_x^{L^2} dy \frac{7 x^2+12 y^2 (-4+\zeta )-2 x y (24+\zeta ) }{ x y^2}G(y)^{2\alpha} Z(y)^{2+2\beta}.
\end{align}
Here the Bose symmetric three-gluon vertex including the renormalization group improvement term from \eref{eq:3g-new} but without IR part is used.
The quadratically divergent parts are those proportional to $\zeta-4$. We deal with them by adding terms to the kernels that cancel them. Since this is a UV problem and we do not want these terms to affect the IR or mid-momentum regime they are multiplied by appropriate damping factors. The kernels become
\begin{align}
 K_{Z}^{gh,\zeta}(p,q) \rightarrow K_{Z}^{gh,\zeta}(p,q)-\frac{1}{12}(4-\zeta)\frac{1}{x y}f_{UV}(y),\\
 K_{Z}^{gl,\zeta}(p,q) \rightarrow K_{Z}^{gl,\zeta}(p,q)+\frac{1}{2}(4-\zeta)\frac{1}{x y}f_{UV}(y),
\end{align}
where
\begin{align}
 f_{UV}(y)=\tanh\left(\frac{y}{L^2}\right).
\end{align}
Using these kernels every loop takes care of its own quadratic divergences. This approach is similar to the one adopted in Ref. \cite{Fischer:2008uz}, where the vertices were modified for this purpose. Since the quadratic divergences are only introduced due to the use of a gauge-variant regularization, they are completely artificial and can consequently be subtracted as done here. However, such a subtraction is not unique and can possibly change the finite parts. Thus we choose the parameter $L$ in an interval for which the results depend least on it.

Next we discuss the parameters $\alpha$ and $\beta$. In order to reproduce the expected UV behavior, they have to fulfill the following condition:
\begin{align}
 2\de=2\alpha\, \de+\gamma\, (2+2\beta).
\end{align}
In order to fix them completely, we demand that this part of the vertex becomes finite in the IR, so that it does not interfere with the IR enhanced part, \eref{eq:3g_IR}. i.e.,
\begin{align}
 \beta=0
\end{align}
for decoupling and 
\begin{align}
 2\beta=\alpha
\end{align}
for scaling.
For the decoupling and scaling solutions we obtain then $\alpha=3+1/\de$, $\beta=0$ and $\alpha=-2-6\de$, $\beta=-1-3\de$, respectively.

The values for $\de$ and $\gamma$ can be obtained by performing the final integral after plugging in the perturbative expressions from eqs. (\ref{eq:G-UV}) and (\ref{eq:Z-UV}) with $s$ chosen such that $G(s)=Z(s)=1$:
\begin{align}
 \left(\omega \log\left(\frac{x}{s}\right)+1 \right)^{-\delta}&= Z_3-\frac{3N_c\,g^2}{64\pi^2(1+\gamma+\delta)\omega}\,\left[\left(1+\omega \log\left(\frac{L^2}{s}\right)\right)^{1+\gamma+\delta}-\left(1+\omega \log\left(\frac{x}{s}\right)\right)^{1+\gamma+\delta}\right]  \\
 \left(\omega \log\left(\frac{x}{s}\right)+1 \right)^{-\gamma}&= \tilde{Z}_3+\frac{\left((\zeta-2)-(24+\zeta)\right)N_c\,g^2}{192\pi^2(1+2\delta)\omega}\,\left[\left(1+\omega \log\left(\frac{L^2}{s}\right)\right)^{1+2\delta}-\left(1+\omega \log\left(\frac{x}{s}\right)\right)^{1+2\delta}\right].
\end{align}
The dependence on $\zeta$ cancels between the two loops. 
The terms divergent for $L\rightarrow \infty$ are canceled by the renormalization functions $Z_3$ and $\tilde{Z}_3$. From the rest the consistency conditions
\begin{align}
 1+\gamma+2\delta&=0,\\
 \frac{3N_c\,g^2}{64\pi^2(1+\gamma+\delta)\omega}&=1,\\
 \frac{13\,N_c\,g^2}{96\pi^2(1+2\delta)\omega}&=1
\end{align}
follow. We obtain for the anomalous dimensions and $\omega$:
\begin{align} 
 \de &=-\frac{9}{44},\\
 \gamma&=-\frac{13}{22},\\
 \omega&= \frac{11\alpha(s)N_c}{12\pi},
\end{align}
with $\alpha(s)=g^2/4\pi$.

\bibliographystyle{utphys_mod}
\bibliography{literature_YM4d}

\end{document}